\documentclass[journal=jctcce,manuscript=article,layout=onecolumn]{achemso}
\setkeys{acs}{articletitle = true}
\pdfoutput=1

\SectionNumbersOn
\mciteErrorOnUnknownfalse

\usepackage[version=3]{mhchem} 
\usepackage[T1]{fontenc}
\usepackage{amsmath}
\usepackage{amssymb}
\usepackage{mathrsfs}

\usepackage{listings}

\usepackage{subcaption}
\usepackage{color}
\usepackage{booktabs}
\usepackage{units}
\usepackage{bm}
\usepackage{braket}
\usepackage{rotating}
\usepackage{mathtools}
\usepackage{appendix}

\newcommand{\ReSpect}{\textsc{ReSpect}}

\DeclareMathOperator{\Tr}{Tr}

\renewcommand{\vec}[1]{\boldsymbol{#1}}

\author{Lukas Konecny}
\affiliation{Hylleraas Centre for Quantum Molecular Sciences, Department of Chemistry, UiT The Arctic University of Norway, N-9037 Troms{\o}, Norway}
\alsoaffiliation{Max Planck Institute for the Structure and Dynamics of Matter, Center for Free Electron Laser Science, Luruper Chaussee 149, 22761 Hamburg, Germany}
\email{lukas.konecny@uit.no}
\author{Stanislav Komorovsky}
\affiliation{Institute of Inorganic Chemistry, Slovak Academy of Sciences, D\'{u}bravsk\'{a} cesta 9, SK-84536 Bratislava, Slovakia}
\author{Jan Vicha}
\affiliation{Centre of Polymer Systems, University Institute, Tomas Bata University in Zl\'{i}n, CZ-76001 Zl\'{i}n, Czech Republic}
\author{Kenneth Ruud}
\affiliation{Hylleraas Centre for Quantum Molecular Sciences, Department of Chemistry, UiT The Arctic University of Norway, N-9037 Troms{\o}, Norway}
\alsoaffiliation{Norwegian Defence Research Establishment, P.O. Box 25, 2027 Kjeller, Norway}
\author{Michal Repisky}
\affiliation{Hylleraas Centre for Quantum Molecular Sciences, Department of Chemistry, UiT The Arctic University of Norway, N-9037 Troms{\o}, Norway}
\alsoaffiliation{Department of Physical and Theoretical Chemistry, Faculty of Natural Sciences, Comenius University, Ilkovicova 6, SK-84215 Bratislava, Slovakia}
\email{michal.repisky@uit.no}

\title[\texttt{achemso} demonstration]
{Exact two-component TDDFT with simple two-electron picture-change corrections:
X-ray absorption spectra near L- and M-edges of four-component quality at two-component cost}

\begin{document}
\begin{abstract}
\noindent

X-ray absorption spectroscopy (XAS) has gained popularity in recent years as it
probes matter with high spatial and elemental sensitivity. However, the
theoretical modelling of XAS is a challenging task since XAS spectra feature a
fine structure due to scalar (SC) and spin--orbit (SO) relativistic effects, in
particular near L and M absorption edges. While full four-component (4c)
calculations of XAS are nowadays feasible, there is still interest in developing
approximate relativistic methods that enable XAS calculations at the
two-component (2c) level while maintaining the accuracy of the parent 4c
approach. In this article we present theoretical and numerical insights into
two simple yet accurate 2c approaches based on an (extended) atomic mean-field
exact two-component Hamiltonian framework, (e)amfX2C, for the calculation of
XAS using linear eigenvalue and damped-response time-dependent density functional theory
(TDDFT). In contrast to the commonly used one-electron X2C (1eX2C) Hamiltonian,
both amfX2C and eamfX2C account for the SC and SO two-electron and
exchange--correlation picture-change (PC) effects that arise from the X2C
transformation. As we demonstrate on L- and M-edge XAS spectra of
transition metal and actinide compounds, the absence of PC corrections in the 1eX2C
approximation results in a substantial overestimatation of SO splittings, whereas (e)amfX2C
Hamiltonians reproduce all essential spectral features such as shape, position,
and SO splitting of the 4c references in excellent agreement, while offering
significant computational savings. Therefore, the (e)amfX2C PC correction models
presented here constitute reliable relativistic 2c 
quantum-chemical approaches for modelling XAS.
\end{abstract}

\section{Introduction}
\label{sec:Introduction}

X-ray absorption spectroscopy (XAS) provides local information about molecular geometry
and electronic structure with high elemental specificity due to the energy separation
of core levels in different elements.~\cite{Besley2020}
XAS spectra are conventionally divided into X-ray absorption fine structure (NEXAFS),
also known as X-ray absorption near-edge structure (XANES), resulting from excitations
of core electrons into vacant bound states, and extended X-ray absorption fine structure
(EXAFS) resulting from the excitations into the continuum.
While the dependence on large-scale experimental facilities
have made X-ray spectroscopy less popular compared to other techniques, advances in instrumentation
are changing this trend. More synchrotrons are being built while at the same time plasma-based
and high harmonic generation-based tabletop sources are making X-ray spectroscopy more accessible
to researches.~\cite{Loh2013,Zhang2016}
We therefore expect an increased interest in XAS and a resulting demand for
fast and reliable theoretical methods to predict and interpret the spectra.

The core electronic states probed by X-rays exhibit relativistic effects, both scalar
relativistic effects presented as constant shifts as well as spin--orbit effects that
split the core p and d levels into \ce{p_{1/2}}--\ce{p_{3/2}}, and \ce{d_{3/2}}--\ce{d_{5/2}}
levels, respectively. These effects are particularly strong in heavy elements such as
transition metals, but are measurable already in XAS spectra of 3rd row elements~\cite{Kadek2015,Vidal2020}.
Therefore, multi-component relativistic quantum chemical methods with variational inclusion
of spin--orbit (SO) coupling present the most reliable approach to these spectral
regions.~\cite{Kasper2020}
The ``gold standard'' in relativistic quantum chemistry is the four-component
(4c) methodology including both scalar and SO effects non-perturbatively via
the one-electron Dirac Hamiltonian in combination with instantaneous Coulomb
interactions among the particles.  Since fully 4c calculations of large
molecular systems containing heavy elements can be time consuming,
researchers have also focused on the development of approximate 2c
Hamiltonians.~\cite{Liu2010,Saue2011} The computational advantage of 2c methods
over 4c methods comes from discarding the negative-energy states, reducing the
original 4c problem by half and also from abandoning the
need to evaluate expensive two-electron integrals over the small-component
basis associated with these states. Examples of popular 2c Hamiltonians are
the second-order Douglas--Kroll--Hess (DKH2)
Hamiltonian,~\cite{Douglas1974,Hess1985,Wolf2002} the zeroth-order regular
approximation (ZORA) Hamiltonian~\cite{Chang1986,Lenthe1993} and the normalized
elimination of small component (NESC) Hamiltonian.~\cite{Dyall1997,Cremer2014}
A 2c Hamiltonian that has gained wide popularity in recent years is the exact
two-component (X2C)
Hamiltonian.~\cite{Heully1986,Jensen2005,Kutzelnigg2005,Liu2007,Ilias2007,Liu2009}
It reduces the 4c problem to 2c by applying simple algebraic manipulations,
avoiding the need to generate explicit operator expressions for (higher-order)
relativistic corrections and/or property operators.

There are several flavours of X2C Hamiltonians differing in the \emph{parent} 4c
Hamiltonian used to construct a 2c model. The use of a pure one-electron (1e)
Dirac Hamiltonian as the parent Hamiltonian results in the one-electron X2C
(1eX2C) where two-electron (2e) interactions are entirely omitted from the X2C
decoupling transformation step~\cite{Saue2011,Konecny2016}. On the other side
is the molecular mean-field X2C approach (mmfX2C)~\cite{Sikkema2009}
where the X2C decoupling is performed \emph{after} converged 4c molecular
self-consistent field (SCF) calculations, making this approach meaningful 
only in connection with post-SCF electron correlation and/or property 
calculations~\cite{Sikkema2009,Halbert2021}. 
Inbetween 1eX2C and mmfX2C there exist several parent Hamiltonian models 
that extend the 1eX2C model by including 2e interactions approximately via 
(i) element and angular-momentum specific screening factors in the evaluation of 
1eSO integrals~\cite{Blume1962,Blume1963};
(ii) a mean-field SO approach~\cite{Hess1996} which has been the basis for
the widely popular AMFI module~\cite{AMFI};
and (iii) an approach that exploits atomic model densities obtained within the 
framework of Kohn--Sham DFT~\cite{Wullen2005,Peng2007}.
The screening factors of type (i) are sometimes referred to as ``Boettger
factors'' or as the screened--nuclear--spin--orbit (SNSO)
approach~\cite{Boettger2000}, obtained from the second-order Douglas--Kroll--Hess 
DFT-based ansatz~\cite{Boettger2000}. Its later reparametrization based on
atomic four-component Dirac--Hartree--Fock results lead to \emph{modified}
SNSO (mSNSO) approach~\cite{Filatov2013}.
Recently, an atomic mean-field (amfX2C) as well as an
extended atomic mean-field (eamfX2C) approach have been presented within the X2C Hamiltonian
framework~\cite{Knecht2022}, extending some of earlier ideas of Liu and Cheng~\cite{Liu2018} 
by comprising the full SO and SC corrections that arise from 2e interactions,
regardless whether they arise from the Coulomb, Coulomb--Gaunt, or
Coulomb--Breit Hamiltonian. Moreover, this ansatz takes into account the
characteristics of the underlying correlation framework, \emph{viz.}, wave-function theory 
or (KS-)DFT, which enables tailor-made exchange--correlation (xc) 
corrections to be introduced~\cite{Knecht2022}.

The 2c approaches are particularly attractive for X-ray spectroscopies where the typical
systems of interest include large transition metal complexes or extended systems such as
surfaces and bulk crystals. This is also reflected in density functional theory (DFT) being
the most popular electronic structure model in XAS calculations since it offers a good
balance between efficiency and accuracy.~\cite{Besley2021}
DFT calculations of XAS spectra can proceed as $\Delta$SCF that subtracts energies of a ground
state and a core-hole excited state~\cite{Bagus1965} or as time-dependent DFT (TDDFT) calculations.
The latter can further be approached in three different ways: real-time TDDFT (RT-TDDFT) propagating
the electronic density in the time domain~\cite{Theilhaber1992,Yabana1996}, the eigenvalue (Casida) equation-based linear response
TDDFT (EV-TDDFT)~\cite{Casida1995,Casida2009}, and damped response TDDFT (DR-TDDFT) (also called the complex
polarization propagator approach) evaluating the spectral function
directly in the frequency domain for the frequencies of interest.~\cite{Oddershede1984,Norman2001,Norman2005}.
RT-TDDFT applied to XAS typically requires simulations with a large number of short time steps
to capture the rapid oscillations associated with core-excited states with sufficient accuracy.
EV-TDDFT yields infinitely-resolved stick spectra in the form of excitation energies
and corresponding transition dipole moments, but for XAS has to rely on core-valence 
separation~\cite{Cederbaum1980, Barth1981, Aagren1994, Aagren1997, Ekstrom2006, Stener2003, Ray2007, Lopata-JCTC-8-3284}
to reach the X-ray spectral region that would otherwise lie too high in the excitation manifold.
On the other hand, DR-TDDFT is an efficient way of targeting the XAS spectral function directly
for user-defined frequencies even in high-frequency and high density-of-states spectral regions
while including relaxation effects and accounting for the finite lifetimes
of the excited states by means of a damping parameter.

TDDFT in the relativistic regime is therefore a perspective approach for
modelling XAS spectra and has already received some attention in the
literature.  Previous XAS applications of 4c-DR-TDDFT have been reported in the
study of the \ce{L3}-edge of \ce{UO2^{2+}}~\cite{South2016}, as well as XAS of
carbon, silicon, germanium, and sulfur compounds.~\cite{Fransson2016_PCCP}.  At the
mSNSO X2C level of theory, these include applications of relativistic EV-TDDFT with
variational SO interactions~\cite{Stetina2019}, nonorthogonal configuration
interaction~\cite{Grofe2022}, and the Bethe--Salpeter equation~\cite{Kehry2020}.

The goal of this article is to provide a careful theoretical formulation of the X2C DR- and EV-TDDFT
starting from the full 4c time-dependent Kohn--Sham (TDKS) equations.
We begin in Section~\ref{sec:Theory} by defining the X2C transformation at the level
of the TDKS equation. We formulate the X2C-transformed time-dependent Fock matrix at the
level of different X2C frameworks, namely the amfX2C, eamfX2C, mmfX2C and 1eX2C.
We introduce the adiabatic X2C approximation where the decoupling matrix is considered
static and independent of the external field, allowing us to derive linear response TDDFT
in the form of damped response and eigenvalue equations. We conclude the Theory section
by summarizing the solvers for DR- and EV-TDDFT and the evaluation of XAS spectra.
Section~\ref{sec:ComputDetails} presents the computational details and
Section~\ref{sec:Results} the results. First, in Section~\ref{sec:Results:Calibration}
a calibration is performed on a series of smaller heavy metal-containing complexes, comparing the
X2C methodologies with reference 4c and experimental data. Then, in Section~\ref{sec:Results:LargerSystems},
amfX2C DR- and EV-TDDFT are showcased on large complexes of chemical interest.
Finally, Section~\ref{sec:Conclusion} provides some concluding remarks.


\section{Theory}
\label{sec:Theory}

Unless stated otherwise, we employ the Einstein summation
convention, SI-based atomic units, and an orthonormal atomic orbital (AO) basis
indexed by two- or four-component flattened subscripts $\mu, \nu, \kappa,
\lambda$.  Each flattened index accounts for scalar, $\tau$, and
multi-component character, $m=2,4$, of relativistic wavefunctions, {\it e.g.} $\mu
\coloneqq m\tau$ (see also Ref.~\citenum{Komorovsky2019}).  Similarly, indices
$i, j$ denote occupied, $a, b$ virtual, and $p, q$ general multicomponent
molecular orbitals (MOs); and subscripts $u, v$ denote Cartesian components. In
addition, we indicate matrix and vector quantities by bold and italic bold
font, respectively.

A convenient starting point for our discussion of DR-TDDFT in an X2C
Hamiltonian framework is to consider the parent four-component (4c) 
equations-of-motion (EOM) for occupied molecular orbital coefficients,
$\boldsymbol{C}^\mathrm{4c}_i(t,\boldsymbol{\mathcal{E}})$. 
Without any loss of generality, we shall consider these equations in the
orthonormal basis
\begin{gather}
  \label{4c:eom}
  i\boldsymbol{\dot{C}}^\mathrm{4c}_i(t,\boldsymbol{\mathcal{E}})
  = 
  \mathbf{F}^\mathrm{4c}(t,\boldsymbol{\mathcal{E}})
  \boldsymbol{C}^\mathrm{4c}_i(t,\boldsymbol{\mathcal{E}}),
\end{gather}
also because our computer implementation generates the corresponding 2c quantities
in such a basis. Eq.~\eqref{4c:eom} describes the molecular system in the
presence of the time-dependent external electric field
$\boldsymbol{\mathcal{E}}(t)$, where the time-dependent Fock matrix written in
either Hartree--Fock or Kohn--Sham theory has the form
\begin{align}
  \label{4c:Fock}
  \mathbf{F}^\mathrm{4c}(t,\boldsymbol{\mathcal{E}})
  = 
  \mathbf{F}_0^\mathrm{4c}\left[\mathbf{D}^\mathrm{4c}(t,\boldsymbol{\mathcal{E}})\right]
  - 
  \mathcal{E}_u(t) \,\mathbf{P}_{\!u}^\mathrm{4c},
\end{align}
with $\mathbf{D}^\mathrm{4c}(t,\boldsymbol{\mathcal{E}})$ being the
time-dependent 4c reduced one-particle density matrix in AO basis
\begin{gather}
  \label{4c:D:t}
  \mathbf{D}^\mathrm{4c}(t,\boldsymbol{\mathcal{E}})
  =
  \boldsymbol{C}^\mathrm{4c}_i(t,\boldsymbol{\mathcal{E}})
  {\boldsymbol{C}^\mathrm{4c}_i}^{\dagger}(t,\boldsymbol{\mathcal{E}}).
\end{gather}
The coupling of a molecular system to a time-dependent external electric field
$\boldsymbol{\mathcal{E}}(t)$---the last term on the right-hand side of
Eq.~\eqref{4c:Fock}---is realized within the dipole approximation by the
electric dipole operator $-\vec{r}$, which is referenced with respect to a gauge
origin $\vec{R}$ and represented in a 4c basis
\begin{gather}
  \label{eq:Poperator}
  P_{u,\mu\nu}^{\mathrm{4c}}
  = 
  -
  \int
  \boldsymbol{X}_{\mu}^{\dagger}(\vec{r})(r_{u}-R_{u})\bm{1}_{4} \boldsymbol{X}_{\nu}(\vec{r})
  d^{3}\vec{r}
  .
\end{gather}
Here, $\bm{1}_{4}$ stands for a $4\!\times\!4$ unit matrix, and the
\emph{orthonormal} four-component $\mu$th AO basis function,
$\boldsymbol{X}_{\mu}(\bm{r}) \coloneqq \boldsymbol{X}^{\text{RKB}}_{\mu}(\bm{r})$,
fulfills the restricted kinetic balance (RKB) condition in its small component
part.~\cite{Stanton1984,ReSpect} Note that the interaction of the system with
the external electric field described here by the dipole operator is an
approximation that can be lifted if necessary.~\cite{List2017, List2020}
However, we do not focus here on this aspect, assuming the
spatial extent of the orbitals involved in the core excitation
being much smaller than the wavelength of the incoming
radiation, which is valid for metal-to-metal transitions
dominating the spectra of the heavy metal complexes considered in the present study.
Also note that we work in the length gauge throughout the paper.
Finally, we assume a generic form of the electric field,
$\boldsymbol{\mathcal{E}}(t)$, in the time domain, because its specific
formulation is only needed later for the discussion related to response theory. 

The field-free Fock matrix $\mathbf{F}_0^\mathrm{4c}$ on the right-hand side of
Eq.~\eqref{4c:Fock} characterizes the molecular system of interest in the
absence of external fields, and consists within the Dirac--Coulomb Hamiltonian
framework of the one-electron Dirac contribution $\mathbf{h}^{\text{4c}}$, the
two-electron contribution, and the exchange--correlation (xc)
contribution
\begin{align}
   \label{eq:Fock_matrix}
   F_{0,\mu\nu}^{\text{4c}}(t,\boldsymbol{\mathcal{E}})
   &=
   h^{\text{4c}}_{\mu\nu}
   +\displaystyle
   G^{\text{4c}}_{\mu\nu,\kappa\lambda}
   D^{\text{4c}}_{\lambda\kappa}(t,\boldsymbol{\mathcal{E}})
   +
   \int v^{xc}_k\!\left[\boldsymbol{\rho}^{\text{4c}}(\vec{r},t,\boldsymbol{\mathcal{E}})\right]
   \Omega_{k,\mu\nu}^{\text{4c}}(\vec{r})\,d^{3}\vec{r}
   ,
\end{align}
with $k=0,\dots,4$. The last term in
Eq.~\eqref{eq:Fock_matrix} is expressed in terms of the
noncollinear xc potential $v_k^{xc}$, that is given within a generalized
gradient approximation (GGA) by
\begin{equation}
   \label{eq:vxc:4c}
   v^{xc}_k\!\left[\boldsymbol{\rho}^{\text{4c}}\right]
   =
   \frac{\partial\varepsilon^{xc}}{\partial \rho_k^{\text{4c}}}
   -
   \left( \boldsymbol{\nabla} \cdot
          \frac{\partial\varepsilon^{xc}}{\partial \boldsymbol{\nabla}\rho_k^{\text{4c}}}\right)
   ,
\end{equation}
where $\varepsilon^{xc}$ refers to a nonrelativistic xc energy density of GGA
type, $\rho_k^{\text{4c}}$
represents the 4c electron charge (for $k=0$) and spin (for $k=1,2,3$)
densities
\begin{equation} 
   \label{eq:ng:4c}
   \rho_k^{\text{4c}} \coloneqq \rho_k^{\text{4c}}(\vec{r},t,\boldsymbol{\mathcal{E}}) 
   = 
   \Omega_{k,\mu\nu}^{\text{4c}}(\vec{r}) D^{\text{4c}}_{\nu\mu}(t,\boldsymbol{\mathcal{E}})
   ,   
\end{equation}
$\boldsymbol{\Omega}_{k}^{\text{4c}}$ are overlap distribution functions
\begin{equation} 
   \label{eq:omega0}
   \Omega^{\text{4c}}_{k,\mu\nu}(\vec{r})
   =
   \boldsymbol{X}_{\mu}^{\dagger}(\vec{r}) \boldsymbol{\Sigma}_k \boldsymbol{X}_{\nu}(\vec{r})
   ,
   \qquad
   \boldsymbol{\Sigma}_{k} =
   \begin{pmatrix}
       \boldsymbol{\sigma}_{k} & \mathbf{0}_2 \\
       \mathbf{0}_2  & \boldsymbol{\sigma}_{k}
   \end{pmatrix}
   , 
\end{equation}
$\boldsymbol{\sigma}_{0} \coloneqq \mathrm{\mathbf{1}}_{2}$,
$(\boldsymbol{\sigma}_1,\boldsymbol{\sigma}_2,\boldsymbol{\sigma}_3)$ is a
vector constructed from the Pauli matrices, and $\mathrm{\mathbf{1}}_{2}$ and
$\mathrm{\mathbf{0}}_{2}$ are $2\times2$ unit and zero matrices, respectively.
In this work we use the noncollinear extension of the nonrelativistic xc
potentials as described in Ref.~\citenum{Komorovsky2019} that is based on the
noncollinear variables of Scalmani and Frisch.\cite{Frisch:2012:non-coll} Note,
however, that because we study closed-shell systems, in the final response
expressions the open-shell $v^{xc}_k$ potential reduces to its closed-shell
form that only depends on the charge density, and to the closed-shell xc kernel
first described in Ref.~\citenum{Radovan-TDDFT-NonColl}.  The round brackets in
Eq.~\eqref{eq:vxc:4c} signify the fact that the gradient operator does not act
on the $\boldsymbol{\Omega}_{k}^{\text{4c}}$ matrix in
Eq.~\eqref{eq:Fock_matrix}. The specific form of the xc contribution to the
Fock matrix in Eqs.~\eqref{eq:Fock_matrix} and~\eqref{eq:vxc:4c} is used here
to simplify the following expressions.  In practical implementations, however,
one uses a formulation that contains $\boldsymbol{\nabla}
\boldsymbol{\Omega}_{k}^{\text{4c}}$ matrices. This formulation can be obtained
from the second term on the RHS of Eq.~\eqref{eq:vxc:4c} by applying partial
integration and employing the fact that both $v^{xc}_k$ and
$\boldsymbol{\Omega}_{k}^{\text{4c}}$ vanish as $|\vec{r}| \rightarrow \infty$.
The two-electron contribution in Eq.~\eqref{eq:Fock_matrix} can be written in
terms of the matrix of generalized anti-symmetrized electron repulsion
integrals (ERIs), 
\begin{equation} 
   \label{eq:eri}
   G^{\text{4c}}_{\mu\nu,\kappa\lambda}
    =
   \mathcal{I}^{\text{4c}}_{\mu\nu,\kappa\lambda}
   -
   \zeta\mathcal{I}^{\text{4c}}_{\mu\lambda,\kappa\nu}
   ;\qquad
   \mathcal{I}^{\text{4c}}_{\mu\nu,\kappa\lambda}
   \coloneqq
   \iint
   \Omega_{0,\mu\nu}^{\text{4c}}(\vec{r}_{1})
   r_{12}^{-1}
   \Omega_{0,\kappa\lambda}^{\text{4c}}(\vec{r}_{2})
   d^{3}\vec{r}_{1}d^{3}\vec{r}_{2}
   ,
\end{equation}
involving the direct and exact-exchange terms, the latter scaled by a scalar
weight factor $\zeta$.  Note that the use of
$\boldsymbol{\Omega}_0^{\text{4c}}(\vec{r})$ in ERIs enables us to write an
efficient relativistic integral algorithm based on complex quaternion
algebra.~\cite{ReSpect} Also note that, similarly to ERIs, there is a dependence
of the xc contribution on the exact-exchange weight factor $\zeta$ for hybrid
functionals. However, we do not write this dependence explicitly to simplify
the notation. 

Before we proceed, let us mention that a convenient way to 
derive DR-TDDFT equations is to employ an ansatz for the
time-dependent MO coefficients~\cite{Konecny2019}
\begin{gather}
  \label{4c:ansatz}
  \boldsymbol{C}^\mathrm{4c}_i(t,\boldsymbol{\mathcal{E}})
  = 
  \boldsymbol{C}^\mathrm{4c}_p d_{pi}(t,\boldsymbol{\mathcal{E}}) e^{-i\varepsilon_it},
\end{gather}
using the reference (static) 4c MO coefficients $\boldsymbol{C}^\mathrm{4c}_p$ as the basis,
and complex-valued $d_{pi}(t,\boldsymbol{\mathcal{E}})$ as expansion coefficients. 
Both the $\boldsymbol{C}^\mathrm{4c}_p$ and the corresponding orbital energies $\varepsilon_p$ 
are obtained from the solution of the time-independent SCF equations~\cite{ReSpect}
\begin{gather}
  \label{4c:Fock:0}
  \mathbf{F}_0^\mathrm{4c}\left[\mathbf{D}^\mathrm{4c}_{0}\right]
  \boldsymbol{C}^\mathrm{4c}_p
  =
  \varepsilon_p \boldsymbol{C}^\mathrm{4c}_p
  ;\qquad
  \mathbf{D}^\mathrm{4c}_{0}
  =
  \boldsymbol{C}^\mathrm{4c}_i
  {\boldsymbol{C}^\mathrm{4c}_i}^{\dagger}.
\end{gather}
Finally, applying this ansatz in the EOM for the 4c MO coefficients
[Eq.~\eqref{4c:eom}], one obtains within linear response theory the 4c
DR-TDDFT expressions as described in Ref.~\citenum{Konecny2019}.

%
%
\noindent
{\bf Transformation of the EOM to the exact two-component (X2C) picture:}
Following the matrix-algebraic approach of X2C, let us assume that for an
arbitrary time and electric field, there exists a unitary transformation matrix
$\mathbf{U}(t,\boldsymbol{\mathcal{E}})$ that block-diagonalizes the 4c Fock
matrix
\begin{gather}
  \label{decoupled_F_C}
  \mathbf{\tilde{F}}^\mathrm{4c}(t,\boldsymbol{\mathcal{E}})
  =
  \mathbf{U}^\dagger(t,\boldsymbol{\mathcal{E}})
  \mathbf{F}^\mathrm{4c}(t,\boldsymbol{\mathcal{E}})
  \mathbf{U}(t,\boldsymbol{\mathcal{E}})
  =
  \left(
    \begin{array}[c]{cc}
      \mathbf{\tilde{F}}^\mathrm{LL}(t,\boldsymbol{\mathcal{E}}) & \bm{0}_{2}      \\
      \bm{0}_{2} & \mathbf{\tilde{F}}^\mathrm{SS}(t,\boldsymbol{\mathcal{E}})      \\
    \end{array}
  \right)
\end{gather}
Note that to be consistent with our previous work, we use a notation with tildes to
indicate picture-change transformed quantities.~\cite{Knecht2022} Under this
transformation, the parent 4c EOM, Eq.~\eqref{4c:eom}, becomes
\begin{gather}
  \label{x4c:eom}
  i \boldsymbol{\dot{\tilde{C}}}^\mathrm{4c}_i(t,\boldsymbol{\mathcal{E}})
  = \mathbf{\tilde{F}}^\mathrm{4c}(t,\boldsymbol{\mathcal{E}})
    \boldsymbol{\tilde{C}}^\mathrm{4c}_i(t,\boldsymbol{\mathcal{E}})
  +
    i \mathbf{\dot{U}}^\dagger(t,\boldsymbol{\mathcal{E}})
    \mathbf{U}(t,\boldsymbol{\mathcal{E}})
    \boldsymbol{\tilde{C}}^\mathrm{4c}_i(t,\boldsymbol{\mathcal{E}}),
\end{gather}
with
\begin{gather} \label{x4c:mos}
  \boldsymbol{\tilde{C}}^\mathrm{4c}_i(t,\boldsymbol{\mathcal{E}})
  =
  \mathbf{U}^\dagger(t,\boldsymbol{\mathcal{E}})
  \boldsymbol{C}^\mathrm{4c}_i(t,\boldsymbol{\mathcal{E}})
  =
  \left(
    \begin{array}[c]{c}
      \boldsymbol{\tilde{C}}^{\mathrm{L}}_{i}(t,\boldsymbol{\mathcal{E}})      \\
      \boldsymbol{\tilde{C}}^{\mathrm{S}}_{i}(t,\boldsymbol{\mathcal{E}})      \\
    \end{array}
  \right).
\end{gather}
Without imposing any additional constrains on the unitary transformation, the
matrix product $\mathbf{\dot{U}}^\dagger(t,\boldsymbol{\mathcal{E}})
\mathbf{U}(t,\boldsymbol{\mathcal{E}})$ has in the general case nonzero
off-diagonal blocks, which prevents a complete decoupling of
Eq.~\eqref{x4c:eom}. However, this term can be safely neglected within the
weak-field and dipole approximations. This statement can be rationalized as
follows. The weak-field approximation, $|\boldsymbol{\mathcal{E}}| \ll 1$, and
the electric dipole approximation, $\omega l c^{-1} \ll 1$, where $l$ refers to
the size of the molecular absorption centre, i.e. a small number for the spatially
localized excitations considered in this work, leads to the estimate
$|\boldsymbol{\mathcal{E}}|\omega l c^{-1} \ll 1$.
It follows that
$\mathbf{\dot{U}}(t,\boldsymbol{\mathcal{E}}) \approx 0$, because the matrix
$\mathbf{\dot{U}}^\dagger(t,\boldsymbol{\mathcal{E}})$ can be estimated as
$O(|\boldsymbol{\mathcal{E}}|\omega c^{-1})$ within the
weak-field approximation (see Ref.~\citenum{Konecny2016}
and Appendix~\ref{appendixA} for a detailed discussion).  As a result, the X2C
transformation is approximately constant in time,
$\mathbf{U}(t,\boldsymbol{\mathcal{E}}) \approx
\mathbf{U}(0,\boldsymbol{\mathcal{E}})$, which is denoted here as the {\it
adiabatic X2C transformation} and its use has already been discussed in the
context of nonlinear optical property calculations in
Ref.~\citenum{Konecny2016}.  In addition, the field-dependence of the X2C
unitary transformation can also be safely neglected within the weak-field
approximation, because the linear-response
$\mathbf{U}^{(1)}(0,\boldsymbol{\mathcal{E}})$ is of order
$|\boldsymbol{\mathcal{E}}|c^{-1}$ (see Appendix~\ref{appendixA}).  As a
consequence of the above discussion,
$\mathbf{\dot{U}}^\dagger(t,\boldsymbol{\mathcal{E}})
\mathbf{U}(t,\boldsymbol{\mathcal{E}}) \approx 0$ and
Eq.~\eqref{x4c:eom} can be written as
\begin{gather}
  \label{x4c:eom:approx}
  i \boldsymbol{\dot{\tilde{C}}}^\mathrm{4c}_i(t,\boldsymbol{\mathcal{E}})
  = \mathbf{\tilde{F}}^\mathrm{4c}(t,\boldsymbol{\mathcal{E}})
    \boldsymbol{\tilde{C}}^\mathrm{4c}_i(t,\boldsymbol{\mathcal{E}}),
\end{gather}
with an X2C unitary transformation that is both time- and electric-field
independent, $\mathbf{U}(t,\boldsymbol{\mathcal{E}}) \approx \mathbf{U}(0,0)$.
In the following discussion we employ the simplified notation, $\mathbf{U}
\coloneqq \mathbf{U}(0,0)$.

The best possible transformation matrix $\mathbf{U}$ which completely decouples
the reference positive-energy MOs $(+)$ from those of negative energy $(-)$ can
be obtained from the so-called mmfX2C approach.~\cite{Sikkema2009} In the
mmfX2C framework, $\mathbf{U} \coloneqq \mathbf{U}_{\text{mmfX2C}}$ is obtained a
posteriori from \emph{converged} time-independent 4c HF/KS equations,
Eq.~\eqref{4c:Fock:0}. As a result, the transformed time-dependent MOs
[Eq.~\eqref{x4c:mos}] become 
\begin{align}
  \label{x4c:ansatz-mmf}
  \boldsymbol{\tilde{C}}^\mathrm{4c}_i(t,\boldsymbol{\mathcal{E}})
  & = 
  \mathbf{U}^\dagger \boldsymbol{C}^\mathrm{4c}_p d_{pi}(t,\boldsymbol{\mathcal{E}}) e^{-i\varepsilon_it}
  =
  \sum_{p\in(+)} 
  \left(
    \begin{array}[c]{c}
      \boldsymbol{\tilde{C}}^\mathrm{L}_{p}  \\
      \boldsymbol{0}                         \\
    \end{array}
  \right)
  d_{pi}(t,\boldsymbol{\mathcal{E}}) e^{-i\varepsilon_it}
  +
  \sum_{p\in(-)} 
  \left(
    \begin{array}[c]{c}
      \boldsymbol{0}                         \\
      \boldsymbol{\tilde{C}}^\mathrm{S}_{p}  \\
    \end{array}
  \right)
  d_{pi}(t,\boldsymbol{\mathcal{E}}) e^{-i\varepsilon_it}
  ,
\end{align}
where 
$\mathbf{\tilde{C}}^\mathrm{L} \coloneqq \mathbf{\tilde{C}}^\mathrm{L}_{\text{mmfX2C}}$
and
$\mathbf{\tilde{C}}^\mathrm{S} \coloneqq \mathbf{\tilde{C}}^\mathrm{S}_{\text{mmfX2C}}$.
Considering the arguments presented in Appendix~\ref{appendixB}, the negative-energy
states [the last term on the RHS of Eq.~\eqref{x4c:ansatz-mmf}] contribute to
the complex polarizability tensor $\bm{\alpha}$ only of order $c^{-4}$. 
By neglecting this contribution, Eq.~\eqref{x4c:eom:approx} becomes
decoupled and one can extract the two-component EOM in the form 
\begin{gather}
  \label{x2c:eom}
  i \boldsymbol{\dot{\tilde{C}}}^\mathrm{2c}_i\!(t,\boldsymbol{\mathcal{E}})
  = 
  \mathbf{\tilde{F}}^\mathrm{2c}(t,\boldsymbol{\mathcal{E}})
  \,\boldsymbol{\tilde{C}}^\mathrm{2c}_i(t,\boldsymbol{\mathcal{E}}).
\end{gather}
Here, both the Fock matrix and occupied positive-energy MO 
coefficients with $i\in(+)$ are picture-change transformed to 2c form as
\begin{align}
   \label{eq:X2Cmmf-F}
   \tilde{F}^{\mathrm{2c}}_{\mu\nu}(t,\boldsymbol{\mathcal{E}})
   &\coloneqq 
   \big(\mathbf{\tilde{F}}^\mathrm{4c}\big)_{\mu\nu}^\mathrm{LL}(t,\boldsymbol{\mathcal{E}})
   =
   \Big[
       \mathbf{U}^{\dagger} \mathbf{F}^{\mathrm{4c}}(t,\boldsymbol{\mathcal{E}}) \mathbf{U}
   \Big]^{\mathrm{LL}}_{\mu\nu}
   ,
   \\
   \label{eq:X2Cmmf-C}
   \tilde{C}^{\mathrm{2c}}_{\mu i}(t,\boldsymbol{\mathcal{E}})
   &\coloneqq 
   \big(\mathbf{\tilde{C}}^\mathrm{4c}\big)_{\mu i}^\mathrm{L}(t,\boldsymbol{\mathcal{E}})
   =
   \Big[ 
        \mathbf{U}^{\dagger} \mathbf{C}^{\mathrm{4c}}(t,\boldsymbol{\mathcal{E}})
   \Big]^{\mathrm{L}}_{\mu i}
   .
\end{align}

%
%
\noindent
{\bf Atomic mean-field X2C (amfX2C):}
As discussed by Knecht~\emph{at al.}\cite{Knecht2022} for the static
time-independent SCF procedure, the correctly transformed 2c Fock matrix
involves the picture-change transformed density matrix, overlap distribution
matrix, as well as one- and two-electron integrals. One may extend this
observation to the time-domain as follows 
\begin{gather}
  \label{eq:2cFock}
  \tilde{F}^{\mathrm{2c}}_{\mu\nu}(t,\boldsymbol{\mathcal{E}})
  =
  \tilde{h}^{\text{2c}}_{\mu\nu}
  +
  \tilde{G}^{\text{2c}}_{\mu\nu,\kappa\lambda}
  \tilde{D}^{\text{2c}}_{\lambda\kappa}(t,\boldsymbol{\mathcal{E}})
  +
  \int v_k^{xc}\!\left[\boldsymbol{\tilde{\rho}}^\text{2c}(\vec{r},t,\boldsymbol{\mathcal{E}})\right]
  \tilde{\Omega}_{k,\mu\nu}^\text{2c}(\vec{r})\,d^{3}\vec{r}
  - 
  \mathcal{E}_u(t) \tilde{P}_{\!u,\mu\nu}^\mathrm{2c}
  .
\end{gather}
The procedure based on Eq.~\eqref{eq:2cFock} leads to results equivalent to the
4c ones, but at a computational cost even higher than its 4c counterpart,
due to the additional picture-change transformation involved.
Therefore, we seek a suitable approximation that enables us to carry out both SCF
iterations as well as linear response calculations in 2c mode such that 
it is computationally efficient and reproduces the reference 4c results as
closely as possible. Keeping this in mind, one can compare
Eq.~\eqref{eq:2cFock} with an approximate and computationally
efficient form of the Fock matrix built with \textit{untransformed}
(without the tilde) two-electron integrals $\mathbf{G}^{\text{2c}}$ and
overlap distribution matrix $\boldsymbol{\Omega}^{\text{2c}}$, that is 
\begin{equation} 
  \label{eq:2cFock_no2ePC}
  F^{\mathrm{2c}}_{\mu\nu}(t,\boldsymbol{\mathcal{E}})
  =
  \tilde{h}^{\text{2c}}_{\mu\nu}
  +
  G^{\text{2c}}_{\mu\nu,\kappa\lambda}
  \tilde{D}^{\text{2c}}_{\lambda\kappa}(t,\boldsymbol{\mathcal{E}})
  +
  \int v_k^{xc}\!\left[\boldsymbol{\rho}^\text{2c}(\vec{r},t,\boldsymbol{\mathcal{E}})\right]
  \Omega_{k,\mu\nu}^\text{2c}(\vec{r})\,d^{3}\vec{r}
  - 
  \mathcal{E}_u(t) \tilde{P}_{\!u,\mu\nu}^\mathrm{2c}
  .
\end{equation}
Here, it is important to emphasize that $\boldsymbol{\rho}^\text{2c}$ also
remains untransformed in the sense that an untransformed
$\boldsymbol{\Omega}_k^{\text{2c}}$ is used but with the correctly transformed density
matrix $\mathbf{\tilde{D}}^{\text{2c}}$. We immediately find
that the difference between these two Fock matrices expresses the
picture-change corrections (PCs) associated with the two-electron integrals and the xc
contribution 
\begin{equation} 
   \label{eq:DeltaF2}
   \Delta\tilde{F}^{\text{2c}}_{\mu\nu}(t,\boldsymbol{\mathcal{E}})
   =
   \tilde{F}^{\text{2c}}_{\mu\nu}(t,\boldsymbol{\mathcal{E}})
   -
   F^{\text{2c}}_{\mu\nu}(t,\boldsymbol{\mathcal{E}})
   =
   \Delta\tilde{G}^{\text{2c}}_{\mu\nu,\kappa\lambda}
   \tilde{D}^{\text{2c}}_{\lambda\kappa}(t,\boldsymbol{\mathcal{E}})
   +
   \Delta\tilde{F}^{\text{2c,xc}}_{\mu\nu}(t,\boldsymbol{\mathcal{E}})
   ,
\end{equation}
where
$
   \Delta\tilde{G}^{\text{2c}}_{\mu\nu,\kappa\lambda}
   = 
   \tilde{G}^{\text{2c}}_{\mu\nu,\kappa\lambda}
   -
   G^{\text{2c}}_{\mu\nu,\kappa\lambda},
$
and
\begin{equation} 
   \Delta\tilde{F}^{\text{2c,xc}}_{\mu\nu}(t,\boldsymbol{\mathcal{E}})
   = 
   \int v_k^{xc}\!\left[\boldsymbol{\tilde{\rho}}^\text{2c}(\vec{r},t,\boldsymbol{\mathcal{E}})\right]
   \tilde{\Omega}_{k,\mu\nu}^{\text{2c}}(\vec{r})\,d^{3}\vec{r}
   -
   \int v_k^{xc}\!\left[\boldsymbol{\rho}^\text{2c}(\vec{r},t,\boldsymbol{\mathcal{E}})\right]
   \Omega_{k,\mu\nu}^{\text{2c}}(\vec{r})\,d^{3}\vec{r}
   .
\end{equation}
In line with the work of Knecht~\emph{et al.},~\cite{Knecht2022} we exploit 
the expected local atomic nature of the differential Fock matrix
$\Delta\tilde{\mathbf{F}}^{\text{2c}}(t,\boldsymbol{\mathcal{E}})$, and
approximate it by a superposition of converged atomic quantities rather
than the converged molecular one, \emph{i.e.}
\begin{align} 
  \label{eq:amfDeltaF2}
  \Delta\mathbf{\tilde{F}}^{\text{2c}}(t,\boldsymbol{\mathcal{E}})
  \approx
  \Delta\mathbf{\tilde{F}}^{\text{2c}}_{\bigoplus}(t,\boldsymbol{\mathcal{E}})
  =
  \bigoplus_{K=1}^{M}
  \Delta\mathbf{\tilde{F}}^{\text{2c}}_{K}[\mathbf{\tilde{D}}_K^{\text{2c}}(t,\boldsymbol{\mathcal{E}})],
\end{align}
where $K$ runs over all atoms in an $M$-atomic system. Such an approach
[Eqs.~\eqref{eq:2cFock_no2ePC}--\eqref{eq:amfDeltaF2}] defines our {\it{atomic
mean-field exact two-component}} (amfX2C) scheme for the two-electron and xc
picture-change corrections applicable to both response and real-time theories.
Note that in contrast to ground-state SCF, where the differential Fock
matrix is governed by transformed atomic density matrices that are
\emph{static} and \emph{perturbation-independent},~\cite{Knecht2022} here the
amfX2C scheme also requires that the \emph{time} and
\emph{perturbation} dependent atomic density matrices are taken into account.  These matrices can be
expanded to first-order in $\boldsymbol{\mathcal{E}}$ as
\begin{align}
  \label{x2c:D-expansion}
  \mathbf{\tilde{D}}_K^\mathrm{2c}(t,\boldsymbol{\mathcal{E}}) 
  =
  \boldsymbol{\tilde{C}}^{\mathrm{2c}}_{K,i}(t,\boldsymbol{\mathcal{E}}) 
  \boldsymbol{\tilde{C}}{^{\mathrm{2c}\,\dagger}_{K,i}}(t,\boldsymbol{\mathcal{E}}) 
  = 
  \mathbf{\tilde{D}}^\mathrm{2c,(0)}_K
  +
  \mathbf{\tilde{D}}^\mathrm{2c,(1)}_{K,u}(t)
  \mathcal{E}_u
  +
  O(|\boldsymbol{\mathcal{E}}|^2)
  ,
\end{align}
where the superscripts $(0)$ and $(1)$ indicate the perturbation-free and
linear-response contributions, respectively [see Eq.~\eqref{eq:def-response}].
The zero- and first-order atomic density matrices
\begin{gather}
  \label{D0:X2C}
  \mathbf{\tilde{D}}^\mathrm{2c}_{K,0}
  \coloneqq
  \mathbf{\tilde{D}}^\mathrm{2c,(0)}_{K}
  =
  \boldsymbol{\tilde{C}}^{\mathrm{2c}}_{K,i}\boldsymbol{\tilde{C}}{^{\mathrm{2c}\,\dagger}_{K,i}}
  ,
  \\
  \label{D1:X2C}
  \mathbf{\tilde{D}}^\mathrm{2c,(1)}_{K,u}(t)
  =
  \boldsymbol{\tilde{C}}_{K,i}^{\mathrm{2c}} \boldsymbol{\tilde{C}}_{K,p}^{\mathrm{2c}\,\dagger}d_{K,u,pi}^{(1)\ast}(t)
  +
  \boldsymbol{\tilde{C}}_{K,p}^{\mathrm{2c}} \boldsymbol{\tilde{C}}_{K,i}^{\mathrm{2c}\,\dagger}d_{K,u,pi}^{(1)}(t)
  ,
\end{gather}
are obtained from the expansion of
$\boldsymbol{\tilde{C}}_{K,i}^\mathrm{2c}(t,\boldsymbol{\mathcal{E}})$ in powers of
$\boldsymbol{\mathcal{E}}$:
\begin{align} 
  \label{x2c:C-expansion}
  \boldsymbol{\tilde{C}}^\mathrm{2c}_{K,i}(t,\boldsymbol{\mathcal{E}})
  & = 
  \boldsymbol{\tilde{C}}^\mathrm{2c}_{K,p} 
  \left[
     \delta_{pi}
     +
     d_{K,u,pi}^{(1)}(t)
     \mathcal{E}_u 
     +
     O(|\boldsymbol{\mathcal{E}}|^2)
  \right]
  e^{-i\varepsilon_{K,i}t}
  .
\end{align}
Here, $d_{K,pi}^{(0)}(t) = \delta_{pi}$ because we have selected as
starting point for the time evolution of
$\boldsymbol{\tilde{C}}^\mathrm{2c}_{K,i}(t,\boldsymbol{\mathcal{E}})$ the
reference atomic orbitals that are eigenvectors of the static Fock matrix
$\mathbf{\tilde{F}}^\mathrm{2c}_{K,0} [\mathbf{\tilde{D}}^\mathrm{2c}_{K,0}]$,
$\boldsymbol{\tilde{C}}^\mathrm{2c}_{K,i}(-\infty,\boldsymbol{\mathcal{E}})
\overset{!}{=} \boldsymbol{\tilde{C}}^\mathrm{2c}_{K,i}$.

Our test calculations reveal that without sacrificing the accuracy of the electric
properties, the evaluation of amfX2C PCs by means of
Eq.~\eqref{eq:amfDeltaF2} can be further simplified by considering only zero-order
atomic density matrices -- that is, by approximating 
\begin{align} 
  \label{eq:amfDeltaF2:static}
  \Delta\mathbf{\tilde{F}}^{\text{2c}}_{\bigoplus}(t,\boldsymbol{\mathcal{E}})
  \approx
  \Delta\mathbf{\tilde{F}}^{\text{amfX2C}}_{\bigoplus}
  =
  \bigoplus_{K=1}^{M}
  \Delta\mathbf{\tilde{F}}^{\text{2c}}_{K}[\mathbf{\tilde{D}}_{K,0}^{\text{2c}}]
  .
\end{align}
This scheme leaves the PCs independent of both time and
$\boldsymbol{\mathcal{E}}$, \emph{i.e.,} it neglects
$O(\boldsymbol{\mathcal{E}})$ terms. In fact, Eq.~\eqref{eq:amfDeltaF2:static}
defines our approximate amfX2C approach for the two-electron and xc
picture-change corrections applicable to both response and real-time theories
involving electric fields, and it is used in the response calculations
reported in this paper. A pseudo-code highlighting the essential steps for evaluating
$\Delta\mathbf{\tilde{F}}^{\text{amfX2C}}_{\bigoplus}$ is available in
Ref.~\citenum{Knecht2022}. With this in mind, the final amfX2C Fock matrix
can be written as 
\begin{align}
  \label{eq:amfX2CFock}
  \tilde{F}^{\mathrm{2c}}_{\mu\nu}(t,\boldsymbol{\mathcal{E}})
  \approx 
  \tilde{F}^{\mathrm{amfX2C}}_{\mu\nu}(t,\boldsymbol{\mathcal{E}})
  & = 
  \tilde{h}^{\text{2c}}_{\mu\nu}
  +
  \Delta\tilde{F}^{\text{amfX2C}}_{\bigoplus,\mu\nu}
  +
  G^{\text{2c}}_{\mu\nu,\kappa\lambda}
  \tilde{D}^{\text{2c}}_{\lambda\kappa}(t,\boldsymbol{\mathcal{E}})
  \nonumber
  \\ 
  & +
  \int v^{xc}_{k}\!\left[\boldsymbol{\rho}^\text{2c}(\vec{r},t,\boldsymbol{\mathcal{E}})\right]
  \Omega_{k,\mu\nu}^{\text{2c}}(\vec{r})\,d^{3}\vec{r}
  - 
  \mathcal{E}_u(t) \tilde{P}_{\!u,\mu\nu}^\mathrm{2c}
  ,
\end{align}
where $\rho_k^\text{2c}(\vec{r},t,\boldsymbol{\mathcal{E}}) = \Omega_{k,\mu\nu}^\text{2c}(\vec{r})
\tilde{D}_{\nu\mu}^\text{2c}(t,\boldsymbol{\mathcal{E}})$ and amfX2C PCs are
represented by the time and perturbation-independent terms
$\Delta\mathbf{\tilde{F}}^{\text{amfX2C}}_{\bigoplus}$.
Note that the decoupling matrix $\mathbf{U}$ in the amfX2C approach is obtained by a
one-step X2C transformation~\cite{Ilias2007,Konecny2016} of the parent 4c
Hamiltonian, $(\mathbf{h}^{\text{4c}} + \mathbf{F}^{\text{4c,2e}}_{\oplus})$,
as defined in Ref.~\citenum{Knecht2022}.

%
%
\noindent
{\bf Extended atomic mean-field X2C (eamfX2C):}
The main advantage of the amfX2C approach is that it introduces picture-change
corrections to both spin-independent and spin-dependent parts of the
two-electron and xc interaction. On the other hand, the fact that these
corrections are only introduced in the atomic diagonal blocks of the
$\Delta\mathbf{\tilde{F}}^{\text{amfX2C}}_{\bigoplus}$ correction means that, for
instance, the direct two-electron contribution will not cancel out with the
electron-nucleus contribution at long distances from the atomic centers. This
becomes problematic in solid-state calculations, where the exact cancellation
of the charge and dipole terms in the expansion at long distances of the direct
two-electron and electron-nucleus contributions is essential. In fact, this was
the main motivation for introducing an extended amfX2C approach (eamfX2C) at
the SCF level.~\cite{Knecht2022}

The generalization of eamfX2C to the time domain requires first to build 
the 4c molecular density matrix $\mathbf{D}^{\text{4c}}_{\bigoplus}(t,\boldsymbol{\mathcal{E}})$ 
and its transformed 2c counterpart 
$\mathbf{\tilde{D}}^{\text{2c}}_{\bigoplus}(t,\boldsymbol{\mathcal{E}})$
from $M$ atomic density matrices $\mathbf{D}^{\text{4c}}_{K}(t,\boldsymbol{\mathcal{E}})$
and $\mathbf{\tilde{D}}^{\text{2c}}_{K}(t,\boldsymbol{\mathcal{E}})$, respectively
\begin{align}
  \label{eq:eamfX2C:D}
  \mathbf{D}^{\text{4c}}_{\bigoplus}(t,\boldsymbol{\mathcal{E}})
  =
  \bigoplus_{K=1}^{M}
  \mathbf{D}^{\text{4c}}_{K}(t,\boldsymbol{\mathcal{E}})
  ,
  \qquad
  \mathbf{\tilde{D}}^{\text{2c}}_{\bigoplus}(t,\boldsymbol{\mathcal{E}})
  =
  \bigoplus_{K=1}^{M}
  \left[
  \mathbf{U}_{K}^{\dagger}(t,\boldsymbol{\mathcal{E}})
  \mathbf{D}^{\text{4c}}_{K}(t,\boldsymbol{\mathcal{E}})
  \mathbf{U}_{K}(t,\boldsymbol{\mathcal{E}})
  \right]^{\text{LL}}
  ,
\end{align}
where all atomic quantities on the RHS of these equations are represented in
\emph{orthonormal} AOs associated with the $K$th atomic center. Then, the
molecular density matrices are used to construct the molecular two-electron
and exchange--correlation Fock contributions in the full molecular basis,
that is
\begin{align}
  \label{eq:F4cplus}
  F^{\text{4c}}_{\bigoplus,\mu\nu}(t,\boldsymbol{\mathcal{E}})
  &=
  G_{\mu\nu,\kappa\lambda}^{\text{4c}}D^{\text{4c}}_{\bigoplus,\lambda\kappa}(t,\boldsymbol{\mathcal{E}}) 
  + 
  \int v_k^{xc}\!\left[\boldsymbol{\rho}^\text{4c}_{\bigoplus}(\vec{r},t,\boldsymbol{\mathcal{E}})\right]
  \Omega_{k,\mu\nu}^{\text{4c}}(\vec{r})\,d^{3}\vec{r}
  ,
  \\
  \label{eq:F2cplus}
  F^{\text{2c}}_{\bigoplus,\mu\nu}(t,\boldsymbol{\mathcal{E}})
  &=
  G_{\mu\nu,\kappa\lambda}^{\text{2c}}\tilde{D}^{\text{2c}}_{\bigoplus,\lambda\kappa}(t,\boldsymbol{\mathcal{E}})
  +
  \int v^{xc}_{k}\!\left[\boldsymbol{\rho}^\text{2c}_{\bigoplus}(\vec{r},t,\boldsymbol{\mathcal{E}})\right]
  \Omega_{k,\mu\nu}^{\text{2c}}(\vec{r})\,d^{3}\vec{r}
  ,
\end{align}
with 
$
\rho_{\bigoplus,k}^\text{4c}(\vec{r},t,\boldsymbol{\mathcal{E}}) 
= 
\Omega_{k,\mu\nu}^{\text{4c}}(\vec{r}) D_{\bigoplus,\nu\mu}^{\text{4c}}(t,\boldsymbol{\mathcal{E}})
$ 
and
$
\rho_{\bigoplus,k}^\text{2c}(\vec{r},t,\boldsymbol{\mathcal{E}}) 
= 
\Omega_{k,\mu\nu}^{\text{2c}}(\vec{r}) \tilde{D}_{\bigoplus,\nu\mu}^{\text{2c}}(t,\boldsymbol{\mathcal{E}})
$.
Transforming the 4c Fock matrix [Eq.~\eqref{eq:F4cplus}] and subtracting the
approximate 2c Fock matrix [Eq.~\eqref{eq:F2cplus}] leads to our
{\it{extended atomic mean-field exact two-component}} (eamfX2C) scheme for
two-electron and xc picture-change corrections applicable to both response and
real-time theories
\begin{align}
  \label{eq:eamfDeltaF2}
  \Delta\mathbf{\tilde{F}}^{\text{2c}}(t,\boldsymbol{\mathcal{E}})
  \approx
  \Delta\mathbf{\tilde{F}}^{\text{2c}}_{\bigoplus}(t,\boldsymbol{\mathcal{E}})
  =
  \left[ \mathbf{U}^{\dagger} \mathbf{F}^{\text{4c}}_{\bigoplus}(t,\boldsymbol{\mathcal{E}}) \mathbf{U} \right]^{\text{LL}}
  -
  \mathbf{F}^{\text{2c}}_{\bigoplus}(t,\boldsymbol{\mathcal{E}})
  .
\end{align}

Following the aforementioned arguments/estimates for the amfX2C approach, the
evaluation of individual atomic density matrices in Eq.~\eqref{eq:eamfX2C:D}
may be approximated in the case of electric properties by their time- and
perturbation-independent components.
This reduces the eamfX2C PCs in Eq.~\eqref{eq:eamfDeltaF2} 
to a significantly simpler form
\begin{align}
  \label{eq:eamfDeltaF2:static}
  \Delta\mathbf{\tilde{F}}^{\text{2c}}_{\bigoplus}(t,\boldsymbol{\mathcal{E}})
  \approx
  \Delta\mathbf{\tilde{F}}^{\text{eamfX2C}}_{\bigoplus}
  =
  \left[ \mathbf{U}^{\dagger} \mathbf{F}^{\text{4c}}_{\bigoplus} \mathbf{U} \right]^{\text{LL}}
  -
  \mathbf{F}^{\text{2c}}_{\bigoplus}
  ,
\end{align}
the evaluation of which can be performed prior to the static SCF as summarized by the
pseudo-code presented in Ref.~\citenum{Knecht2022}. The final picture-change
transformed 2c Fock matrix with PCs represented by
Eq.~\eqref{eq:eamfDeltaF2:static} defines our approximate eamfX2C approach
suitable for both response and real-time theories
\begin{align}
  \label{eq:eamfX2CFock}
  \tilde{F}^{\mathrm{2c}}_{\mu\nu}(t,\boldsymbol{\mathcal{E}})
  \approx 
  \tilde{F}^{\mathrm{eamfX2C}}_{\mu\nu}(t,\boldsymbol{\mathcal{E}})
  & = 
  \tilde{h}^{\text{2c}}_{\mu\nu}
  +
  \Delta\tilde{F}^{\text{eamfX2C}}_{\bigoplus,\mu\nu}
  +
  G^{\text{2c}}_{\mu\nu,\kappa\lambda}
  \tilde{D}^{\text{2c}}_{\lambda\kappa}(t,\boldsymbol{\mathcal{E}})
  \nonumber
  \\ 
  & +
  \int v_k^{xc}\!\left[\boldsymbol{\rho}^\text{2c}(\vec{r},t,\boldsymbol{\mathcal{E}})\right]
  \,\Omega_{k,\mu\nu}^\text{2c}(\vec{r})\,d^{3}\vec{r}
  - 
  \mathcal{E}_u(t) \tilde{P}_{\!u,\mu\nu}^\mathrm{2c}
  ,
\end{align}
with $\rho_k^\text{2c}(\vec{r},t,\boldsymbol{\mathcal{E}}) = \Omega_{k,\mu\nu}^\text{2c}(\vec{r})
\tilde{D}_{\nu\mu}^\text{2c}(t,\boldsymbol{\mathcal{E}})$.
Note that the decoupling matrix $\mathbf{U}$ in the eamfX2C approach is obtained by a
one-step X2C transformation~\cite{Ilias2007,Konecny2016} of the parent 4c
Hamiltonian, $(\mathbf{h}^{\text{4c}} + \mathbf{F}^{\text{4c,2e}}_{\oplus})$,
as defined in Ref.~\citenum{Knecht2022}.

%
%
\noindent
{\bf One-electron X2C (1eX2C) and molecular mean-field X2C (mmfX2C):}
The above-mentioned amfX2C and eamfX2C Hamiltonian models for response and
real-time theories lie in between two extreme cases. The first one is represented
by a pure one-electron X2C (1eX2C) Hamiltonian where two-electron as well as xc
picture-change corrections are omitted entirely. The resulting 1eX2C Fock
matrix then reads 
\begin{align}
  \label{eq:1eX2CFock}
  \tilde{F}^{\mathrm{2c}}_{\mu\nu}(t,\boldsymbol{\mathcal{E}})
  \approx 
  \tilde{F}^{\mathrm{1eX2C}}_{\mu\nu}(t,\boldsymbol{\mathcal{E}})
  & = 
  \tilde{h}^{\text{2c}}_{\mu\nu}
  +
  G^{\text{2c}}_{\mu\nu,\kappa\lambda}
  \tilde{D}^{\text{2c}}_{\lambda\kappa}(t,\boldsymbol{\mathcal{E}})
  \nonumber
  \\ 
  & +
  \int v_k^{xc}\!\left[\boldsymbol{\rho}^\text{2c}(\vec{r},t,\boldsymbol{\mathcal{E}})\right]
  \,\Omega_{k,\mu\nu}^\text{2c}(\vec{r})\,d^{3}\vec{r}
  - 
  \mathcal{E}_u(t) \tilde{P}_{\!u,\mu\nu}^\mathrm{2c}
  ,
\end{align}
where the decoupling matrix $\mathbf{U}$ is obtained simply from the parent
one-electron Dirac Hamiltonian~\cite{Ilias2007,Konecny2016}. Due to its
simplicity the 1eX2C Hamiltonian still remains very popular, but caution is
needed when applying this model beyond valence electric properties as shown in
this article. The second model, coined as molecular mean-field X2C
(mmfX2C)~\cite{Sikkema2009}, requires to perform a full molecular 4c SCF
calculation in order to determine $\mathbf{U}$ from converged 4c solutions.
The subsequent X2C transformation of the converged 4c Fock matrix gives the
mmfX2C Fock matrix:
\begin{align}
   \label{eq:SCFmmfX2CFock}
   \tilde{F}^{\mathrm{mmfX2C}}_{\mu\nu}
   & = 
   \tilde{h}^{\text{2c}}_{\mu\nu}
   +
   \Delta\tilde{F}^{\text{mmfX2C}}_{\mu\nu}
   +
   G^{\text{2c}}_{\mu\nu,\kappa\lambda}
   \tilde{D}^{\text{2c}}_{\lambda\kappa}
   +
   \int v_k^{xc}\!\left[\boldsymbol{\rho}^\text{2c}\right]
   \,\Omega_{k,\mu\nu}^\text{2c}(\vec{r})\,d^{3}\vec{r}
   ,
\end{align}
eigenvalues of which agree up to the computer precision with eigenvalues of
positive-energy MOs of the parent 4c Fock matrix. Note that in
Eq.~\eqref{eq:SCFmmfX2CFock}, the two-electron and xc picture-change
corrections are included exactly through the differential Fock matrix
$\Delta\mathbf{\tilde{F}}^{\text{mmfX2C}}$ evaluated according to
Eq.~\eqref{eq:DeltaF2} (or more precisely by its time- and
perturbation-independent counterpart). By following the arguments that justify
the use of time- and perturbation-independent (e)amfX2C picture-change
correction models in response and real-time theories involving electric
field(s) [Eqs.~\eqref{eq:amfX2CFock} and~\eqref{eq:eamfX2CFock}], one can also
define a similar model for mmfX2C with the Fock matrix
\begin{align}
  \label{eq:mmfX2CFock}
  \tilde{F}^{\mathrm{2c}}_{\mu\nu}(t,\boldsymbol{\mathcal{E}})
  \approx 
  \tilde{F}^{\mathrm{mmfX2C}}_{\mu\nu}(t,\boldsymbol{\mathcal{E}})
  & = 
  \tilde{h}^{\text{2c}}_{\mu\nu}
  +
  \Delta\tilde{F}^{\text{mmfX2C}}_{\mu\nu}
  +
  G^{\text{2c}}_{\mu\nu,\kappa\lambda}
  \tilde{D}^{\text{2c}}_{\lambda\kappa}(t,\boldsymbol{\mathcal{E}})
  \nonumber
  \\ 
  & +
  \int v_k^{xc}\!\left[\boldsymbol{\rho}^\text{2c}(\vec{r},t,\boldsymbol{\mathcal{E}})\right]
  \,\Omega_{k,\mu\nu}^\text{2c}(\vec{r})\,d^{3}\vec{r}
  - 
  \mathcal{E}_u(t) \tilde{P}_{\!u,\mu\nu}^\mathrm{2c}
  .
\end{align}
Here, $\rho_k^\text{2c}(\vec{r},t,\boldsymbol{\mathcal{E}}) = \Omega_{k,\mu\nu}^\text{2c}(\vec{r})
\tilde{D}_{\nu\mu}^\text{2c}(t,\boldsymbol{\mathcal{E}})$.

%
%
\noindent
{\bf Linear DR-TDDFT within the presented X2C approaches:}
Before we proceed, let us specify the form of the time-dependent electric field
$\mathcal{E}_u(t)$ in Eqs.~\eqref{eq:amfX2CFock} and~\eqref{eq:eamfX2CFock}. In
response theory, it is customary to choose the electric field to have the
form of a harmonic field of frequency $\omega$ and amplitude
$2\boldsymbol{\mathcal{E}}$ that is slowly switched on using the factor $\eta$,
that is
\begin{align}
  \label{eq:Et}
  \mathcal{E}_u(t) = \mathcal{E}_u \mathcal{F}(t),
  \qquad
  \mathcal{F}(t) = \left( e^{-i\omega t+\eta t} + e^{i\omega t+\eta t}\right)
  .
\end{align}
Note, that the purpose of the field-switching factor $\eta$ is to ensure a
smooth application of the electric field and the limit $\eta \rightarrow 0$ is
considered later in the derivation. This is in contrast to the damping
factor $\gamma$, which describes the rate of the relaxation of the system and
should enter the RHS of the EOM in a separate term. However, such a parameter
can only enter a relaxation-including equation such as the Liouville--von Neumann
equation, an EOM for the density matrix. To simplify the discussion above, we have
omitted the factor $\gamma$ altogether and worked with an EOM for the MO coefficients.
Later we add this factor {\it ad hoc} in Eq.~\eqref{eq:d1-dcc-difeq}, and we do not
discuss the factor $\eta$ beyond this paragraph. We refer the interested reader to a more
detailed discussion of the factors $\eta$ and $\gamma$ in
Refs.~\citenum{NormanRuudSaue, Konecny2019}.  

While a direct time propagation of the 2c EOM, Eq.~\eqref{x2c:eom},
results in the 2c-RT-TDHF or 2c-RT-TDDFT approaches,\cite{Konecny2016} response
theory rather seeks for its solution via a perturbation expansion.
To this end, we write the expansion of the 2c MO
coefficients $\boldsymbol{\tilde{C}}^\mathrm{2c}_{i}(t,\boldsymbol{\mathcal{E}})$
in powers of the external field
\begin{align} 
  \label{eq:C2c-epansion}
  \boldsymbol{\tilde{C}}^\mathrm{2c}_{i}(t,\boldsymbol{\mathcal{E}})
  & = 
  \sum_{p\in(+)}
  \boldsymbol{\tilde{C}}^\mathrm{L}_{p} 
  \left[
     \delta_{pi}
     +
     d_{u,pi}^{(1)}(t)
     \mathcal{E}_u 
     +
     O(|\boldsymbol{\mathcal{E}}|^2)
  \right]
  e^{-i\varepsilon_{i}t}
  ,
\end{align}
where $d^{(1)}_{u,pi}(t)$ are the first-order expansion coefficients whose
Fourier components are in the end determined within the linear-response regime.
In Eq.~\eqref{eq:C2c-epansion}, we have neglected negative-energy states, as discussed below
Eq.~\eqref{x4c:ansatz-mmf} and in Appendix~\ref{appendixB}. In addition, we
assume
$\boldsymbol{\tilde{C}}^\mathrm{2c}_{i}(-\infty,\boldsymbol{\mathcal{E}})
\overset{!}{=} \boldsymbol{\tilde{C}}^\mathrm{2c}_{i}$ which leads to
$d_{pi}^{(0)}(t) = \delta_{pi}$. The reference 2c MOs
$\boldsymbol{\tilde{C}}^\mathrm{2c}_{i}$ and one-electron energies
$\varepsilon_{i}$ are eigenvectors and eigenvalues of the static Fock matrix
$\mathbf{\tilde{F}}^\mathrm{2c}_0[\mathbf{\tilde{D}}^\mathrm{2c}_{0}] \coloneqq
\mathbf{\tilde{F}}^\mathrm{2c}(-\infty,0)$ 
[see Eqs.~\eqref{eq:amfX2CFock}, \eqref{eq:eamfX2CFock}, \eqref{eq:1eX2CFock}, 
and~\eqref{eq:mmfX2CFock}], respectively. Applying the ansatz in
Eq.~\eqref{eq:C2c-epansion} to the EOM in Eq.~\eqref{x2c:eom} one can extract the
differential equation for the first-order perturbation coefficients
$\mathbf{d}_u^{(1)}$ as follows
\begin{align}
\label{eq:d1-dcc-difeq}
i \frac{d}{dt}
\begin{pmatrix}
\mathbf{d}^{(1)}_u(t) \\
\mathbf{d}^{(1)\ast}_u(t)
\end{pmatrix}
= &
\begin{pmatrix}
 \mathbf{A}^{\!\text{2c}} - i\gamma\bm{1}  &  \mathbf{B}^\text{2c} \\
-\mathbf{B}^{\text{2c}\ast} & -\mathbf{A}^{\!\text{2c}\ast} - i \gamma\bm{1}
\end{pmatrix}
\begin{pmatrix}
\mathbf{d}^{(1)}_u(t) \\
\mathbf{d}^{(1)\ast}_u(t)
\end{pmatrix}
-
\begin{pmatrix}
 \mathbf{\tilde{P}}_u^\text{2c} \mathcal{F}(t)
 \\
-\mathbf{\tilde{P}}_u^{\text{2c}\ast} \mathcal{F}(t)
\end{pmatrix}
,
\end{align}
where $\mathbf{d}^{(1)}$ and $\mathbf{P}$ are complex matrices of size
$N_\mathrm{v} \times N_\mathrm{o}$ with $N_\mathrm{v}$ and $N_\mathrm{o}$
referring to the number of virtual and occupied MOs, respectively.
Eq.~\eqref{eq:d1-dcc-difeq} is written in terms of the virtual--occupied coefficients
$d_{u,ai}^{(1)}$ because the occupied--occupied and virtual--virtual ones do not contribute to the
time-dependent density matrix (see Ref.~\citenum{Konecny2019} for further
details). In Eq.~\eqref{eq:d1-dcc-difeq} the matrices
$\mathbf{A}^\text{2c}$ and $\mathbf{B}^\text{2c}$ are defined
as
\begin{align}
  \label{eq:Aterm}
  A^\text{2c}_{ai,bj} &= \omega_{ai}\delta_{ab}\delta_{ij} +
       \left( G^\text{2c}_{\mu\nu,\kappa\lambda} + K^\mathrm{xc}_{\mu\nu,\kappa\lambda} \right)
       \tilde{C}^{\text{L}\ast}_{\mu a} \tilde{C}^\text{L}_{\nu i}
       \tilde{C}^{\text{L}\ast}_{\kappa j} \tilde{C}^\text{L}_{\lambda b}
  ,
  \\
  \label{eq:Bterm}
  B^\text{2c}_{ai,bj} &=
       \left( G^\text{2c}_{\mu\nu,\kappa\lambda} + K^\mathrm{xc}_{\mu\nu,\kappa\lambda} \right)
       \tilde{C}^{\text{L}\ast}_{\mu a} \tilde{C}^\text{L}_{\nu i}
       \tilde{C}^{\text{L}\ast}_{\kappa b} \tilde{C}^\text{L}_{\lambda j}
  ,
  \\
  \label{eq:Kterm}
  \mathbf{K}^\mathrm{xc} &= \mathbf{K}^\mathrm{xc}
       \Big( \boldsymbol{\Omega}_{k}^\text{2c}, \mathbf{\tilde{D}}^\text{2c} \Big)
  ,
\end{align}
where no summation is assumed in the first term on the RHS of
Eq.~\eqref{eq:Aterm}, and $\omega_{ai} = \varepsilon_a - \varepsilon_i$ with
$\varepsilon_p$ being the one-electron energy of the $p$th molecular orbital.
In Eqs.~\eqref{eq:Aterm}--\eqref{eq:Kterm}, $\mathbf{G}^\text{2c}$ are the 2c
untransformed two-electron integrals, and $\mathbf{K}^\mathrm{xc}$ is the
exchange--correlation kernel constructed from 2c untransformed overlap
distribution functions $\boldsymbol{\Omega}_{k}^\text{2c}$ and the transformed 2c
density matrix $\mathbf{\tilde{D}}^\text{2c}$. The functional form of the
noncollinear xc kernel follows the one presented by Bast~{\emph et
al.}\cite{Radovan-TDDFT-NonColl}, however in this work we utilize the RKB basis
in contrast to the unrestricted kinetic balance basis employed in
Ref.~\citenum{Radovan-TDDFT-NonColl}.

The differential equation in Eq.~\eqref{eq:d1-dcc-difeq} can be turned into an
algebraic form by the method of undetermined coefficients, substituting
\begin{equation}
  \label{eq:rspXYansatz}
  \mathbf{d}_u^{(1)}(t)
  = 
  \mathbf{X}_u e^{- i \omega t + \gamma t} + \mathbf{Y}_u^\ast e^{i \omega t + \gamma t}
  ,
\end{equation}
where $\mathbf{X}_u$ and $\mathbf{Y}_u$ are complex matrices of time-independent undetermined
coefficients. After substituting Eq.~\eqref{eq:rspXYansatz} into Eq.~\eqref{eq:d1-dcc-difeq}
and collecting terms proportional to $e^{-i \omega t + \gamma t}$, one arrives at the final
linear damped response equation
\begin{equation}
  \label{eq:linearDampedResponse}
  \left[
  \begin{pmatrix}
  \mathbf{A}^{\!\text{2c}}   & \mathbf{B}^\text{2c} \\
  \mathbf{B}^{\text{2c}\ast} & \mathbf{A}^{\!\text{2c}\ast}
  \end{pmatrix}
  - (\omega+i\gamma)
  \begin{pmatrix}
  \mathbf{1} & \mathbf{0} \\
  \mathbf{0} & -\mathbf{1}
  \end{pmatrix}
  \right]
  \begin{pmatrix}
  \mathbf{X}_{u} \\
  \mathbf{Y}_{u}
  \end{pmatrix}
  =
  \begin{pmatrix}
  \mathbf{\tilde{P}}^\mathrm{2c}_{u} \\
  \mathbf{\tilde{P}}^{\mathrm{2c}\ast}_{u}
  \end{pmatrix}
  .
\end{equation}
Both, $\omega$ and $\gamma$ are user-defined parameters specifying the external
electric field frequency and a common relaxation (damping) parameter modelling
the finite lifetime of the excited states that leads to finite-width peaks. The
right-hand side of Eq.~\eqref{eq:linearDampedResponse} describes the
interaction of the molecular system with the applied external electric field,
which in the electric dipole approximation is mediated by the electric dipole
moment operator.

In addition, the solution of the homogeneous form of the linear system of
differential equations in Eq.~\eqref{eq:d1-dcc-difeq}, {\it i.e.} for
$\boldsymbol{\mathcal{E}}=0$, leads to the eigenvalue TDDFT (EV-TDDFT) equation
\begin{equation}
  \label{eq:eigenvalueTDDFT}
  \begin{pmatrix}
  \mathbf{A}^{\!\text{2c}}   & \mathbf{B}^\text{2c} \\
  \mathbf{B}^{\text{2c}\ast} & \mathbf{A}^{\!\text{2c}\ast}
  \end{pmatrix}
  \begin{pmatrix}
  \mathbf{X}_{N} \\
  \mathbf{Y}_{N}
  \end{pmatrix}
  =
  \omega_N
  \begin{pmatrix}
  \mathbf{1} & \mathbf{0} \\
  \mathbf{0} & -\mathbf{1}
  \end{pmatrix}
  \begin{pmatrix}
  \mathbf{X}_{N} \\
  \mathbf{Y}_{N}
  \end{pmatrix}
  ,
\end{equation}
where $\omega_N$ represents a vertical electronic excitation energy from the
reference state to the $N$-th excited state with a transition vector
$(\mathbf{X}_{N} \mathbf{Y}_{N})^\mathrm{T}$. This equation represents
another linear-response TDDFT approach to molecular properties such as XAS spectra.
Note that despite the similarity in notation, the response vector $(\mathbf{X}_u \mathbf{Y}_u)^\mathrm{T}$
of the damped response TDDFT and the transition vector $(\mathbf{X}_{N} \mathbf{Y}_{N})^\mathrm{T}$
of the eigenvalue TDDFT have different meaning and units. While the former describes the response
to an external perturbation and depends on its operator, frequency, and damping factor, the latter
only describes the transition amplitude between the ground state and $N$-th excited state.

DR-TDDFT, Eq.~\eqref{eq:linearDampedResponse}, is solved using an
iterative subspace algorithm, since the size of the matrix on the left-hand side of the equation
prohibits its direct inversion or the use of elimination techniques for realistic molecular systems.
Because the equation has the same properties in terms of complexity and symmetries as the
4c DR-TDDFT equation, we can employ the same solver as is used for the 4c case.~\cite{Villaume2010, Konecny2019}
The iterative subspace  solver implemented  in the \ReSpect{} program~\cite{ReSpect}
explicitly treats the terms in the response equation based on their hermicity and time-reversal
symmetry, and allows several frequencies (tens to hundreds) to be considered simultaneously,
thus covering a large part of the spectrum in a single run.
A detailed presentation of this solver is available in Ref.~\citenum{Konecny2019}.
Similarly, the eigenvalue linear response TDDFT equation, Eq.~\eqref{eq:eigenvalueTDDFT},
is also solved iteratively by a variant of the Davidson--Olsen algorithm, as presented
in Ref.~\citenum{Komorovsky2019}.

The calculation of XAS spectra in the linear response regime corresponds to evaluating
the dipole strength function
\begin{equation}
\label{eq:rspFunction}
S(\omega)
=
\frac{4\pi\omega}{3 c} \Im \Tr \left[ \bm{\alpha}(\omega) \right]
,
\end{equation}
where $c$ is the speed of light, $\Im$ denotes the imaginary part, Tr the trace over the
Cartesian components, and $\bm{\alpha}(\omega)$ is the complex polarizability tensor
in the frequency domain. This tensor parametrizes the first-order electric dipole response,
i.e. the induced electric dipole moment $\bm{\mu}^\mathrm{ind}(\omega)$, to an external electric
field as 
\begin{equation}
\mu^\mathrm{ind}_u (\omega)
=
\alpha_{uv} (\omega) E_v (\omega) + \ldots
.
\end{equation}
In 2c-DR-TDDFT, the $\bm{\alpha}(\omega)$ tensor components are calculated for a
user-defined set of frequencies from the response vector $(\mathbf{X}_u \mathbf{Y}_u)^\mathrm{T}$,
the solution of Eq.~\eqref{eq:linearDampedResponse}, via
\begin{equation}
\label{eq:polarisability-DR}
\alpha_{uv}(\omega)
=
X_{ai,v}(\omega) \tilde{P}^{\mathrm{2c}}_{ia,u} + Y_{ai,v}(\omega) \tilde{P}^{\mathrm{2c}}_{ai,u}
.
\end{equation}
In contrast, the evaluation of the complex polarizability tensor from the solution $(\mathbf{X}_{N} \mathbf{Y}_{N})^\mathrm{T}$
of the EV-TDDFT equation, Eq.~\eqref{eq:eigenvalueTDDFT}, proceeds via a calculation of the transition dipole moment
\begin{equation}
\label{eq:transitionDipoleEV}
t_{u,N}
=
X_{ai,N}(\omega) \tilde{P}^{\mathrm{2c}}_{ia,u} + Y_{ai,N}(\omega) \tilde{P}^{\mathrm{2c}}_{ai,u}
,
\end{equation}
which is then inserted into the expression for the polarizability as a linear response function
\begin{equation}
\label{eq:polarisability-EV}
\alpha_{uv}(\omega)
=
\sum_N
\left[
\frac{t_{u,N}^* t_{v,N}}{\omega + \omega_N + i\gamma}
-
\frac{t_{u,N} t_{v,N}^*}{\omega - \omega_N + i\gamma}
\right]
.
\end{equation}
Here, the frequency $\omega$ and damping parameter $\gamma$ are included in the calculation
of $\alpha_{uv}(\omega)$, i.e. essentially in a post-processing step, while in DR-TDDFT they
are terms in the main working equation, Eq.~\eqref{eq:linearDampedResponse}.
This difference has consequences for the workflow and practicality of the DR-TDDFT and eigenvalue
TDDFT in various situations, even though the methods yield identical final spectra
assuming the same $\gamma$ factor is used.


\section{Computational details}
\label{sec:ComputDetails}

For the purpose of benchmarking and calibration, a set of closed-shell heavy metal-containing compounds
with high-quality experimental data available, including 3d, 4d, 5d, and 5f elements with various
electron configurations of the central atom was selected, specifically
\ce{VOCl3}, \ce{CrO2Cl2}, \ce{MoS4^{2-}}, \ce{WCl6}, \ce{PdCl6^{2-}}, \ce{ReO4-}, and \ce{UO2(NO3)2}.
Moreover, XAS spectra of larger systems, namely \ce{[Ru Cl2 (DMSO)2 (Im)2]} (Im = imidazole, DMSO = dimethyl
sulfoxide), \ce{[W Cl4 (PMePh2)2]} (Ph = phenyl), and \ce{[(\eta^{6}-p-cym)Os(Azpy-NMe2)I]^{+}}
(p-cym = p-cymene, \ce{Azpy-NMe 2} = 2-(p-([dimethylamino]phenylazo)pyridine)))
are included.  
Geometries were optimized using \textsc{TURBOMOLE} quantum-chemical program~\cite{Balasubramani2020}
with a protocol designed for transition metal
elements~\cite{Vicha2013, Vicha2015}:
PBE0 functional~\cite{Perdew1996, Perdew1997, Adamo1999, Slater1951},
def2-TZVPP basis sets~\cite{Weigend2005}
for all atoms (def-TZVP for uranium complex)
with corresponding effective core potentials (ECPs)~\cite{Andrae1990}
replacing 28 core electrons in 4d and 60 electrons in 5d and 5f elements.

All X-ray spectra were calculated using the damped response library~\cite{Konecny2019}
and linear response TDDFT library~\cite{Komorovsky2019}
of the Relativistic Spectroscopy DFT program \ReSpect{}.\cite{ReSpect}
Uncontracted all-electron GTO basis sets were used for all systems.
The selected basis sets
were the uncontracted Dyall’s VDZ basis sets~\cite{Dyall2007-4d, Dyall2004-5d, Dyall2010-5d-rev,
Dyall2007-5f,Dyall1998-4p-6p, Dyall2006-4p-6p-rev} for metals and iodine (basis sets for 3d elements
are available upon request) and the uncontracted
Dunning’s aug-cc-pVDZ basis sets~\cite{Dunning1989, Kendall1992, Woon1993} for light elements.
The systems were treated using the PBE0 density functional including a modified
version PBE0-$x$HF with variable exact-exchange admixture $x$ that was previously
shown to be crucial to counter the shifts observed with standard parametrizations.~\cite{Konecny2022}
The numerical integration of the noncollinear exchange-correlation potential and kernel was done with an adaptive
molecular grid of medium size (program default).
In the 2c calculations, atomic nuclei of finite size were approximated by a Gaussian
charge distribution model.\cite{Visscher1997}

The damped linear response calculations covered the spectral regions with a resolution of
\unit[0.1]{eV}. The initial guess of the spectral regions to be scanned were provided
by the orbital energies of the target core orbitals. Core-valence
separation~\cite{Cederbaum1980, Barth1981, Aagren1994, Aagren1997, Ekstrom2006, Stener2003, Ray2007, Lopata-JCTC-8-3284}
was used to remove non-physical valence-to-continuum excitations that may occur at the same energy ranges
as the physical core excitations.
All damped response calculations employed the multi-frequency solver with 100 frequencies
treated simultaneously.
The damping/broadening parameter used in the damped response calculations was set to \unit[0.15]{eV}
for high-resolution spectra, while values ranging from 0.5 to \unit[3]{eV} were used to obtain wider
peaks to facilitate the comparison with experimental line shapes.
Since the value of the damping parameter $\gamma$ affects the amplitude of the spectra, in graphs where
we compare spectra with different damping parameters or with normalized experimental spectra, we normalize
the calculated spectra to unity. This is denoted by arbitrary units (arb. units) instead of atomic units (au)
as the dimension of the spectral function.

The eigenvalue linear response TDDFT calculations used core-valence separation to access excitation
energies associated with core-excited states. The eigenvalue equation was solved iteratively for the
first 50 excitation energies. The spectra were subsequently calculated from the excitation energies
and transition moments obtained from the eigenvectors with the same Lorentzian broadening as in the
corresponding DR-TDDFT calculations.


\section{Results and discussion}
\label{sec:Results}

\subsection{Calibration of X2C approaches}
\label{sec:Results:Calibration}

\begin{figure}[ht]
\caption{
amfX2C Hamiltonian reproduces the 4c reference while 1eX2C Hamiltonian overestimates
spin--orbit splitting leading to incorrect 1eX2C spectral shape
of XAS spectra of \ce{MoS4^{2-}} near Mo \ce{M_{4,5}}-edges calculated using PBE0 functional
and DZ/aDZ basis sets.
}
\label{fig:MoS4-X2Cvs4C}
\centering
\includegraphics[width=0.65\textwidth]{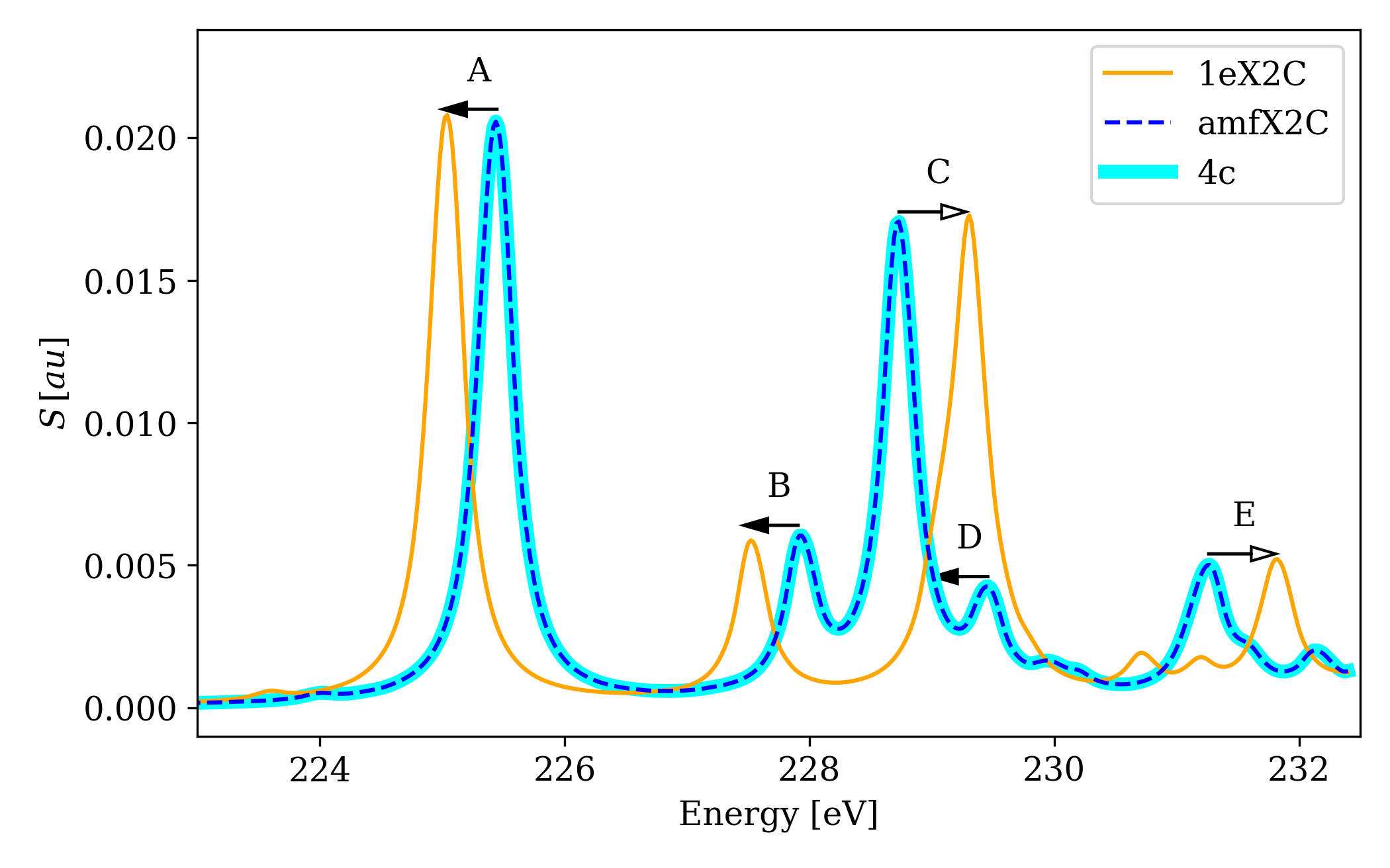}
\end{figure}

\begin{figure}[ht]
\caption{
amfX2C Hamiltonian reproduces the 4c reference while 1eX2C Hamiltonian overestimates
spin--orbit splitting leading to \unit[20]{eV} shifts 
in XAS spectra of \ce{WCl6} near W \ce{L_{2,3}}-edges calculated using PBE0-60HF functional
and DZ/aDZ basis sets. Broad peaks were obtained with damping parameter $\gamma=\unit[3.0]{eV}$
while narrow ones with $\gamma=\unit[0.15]{eV}$.
}
\label{fig:WCl6-X2C-SO}
\centering
\includegraphics[width=0.65\textwidth]{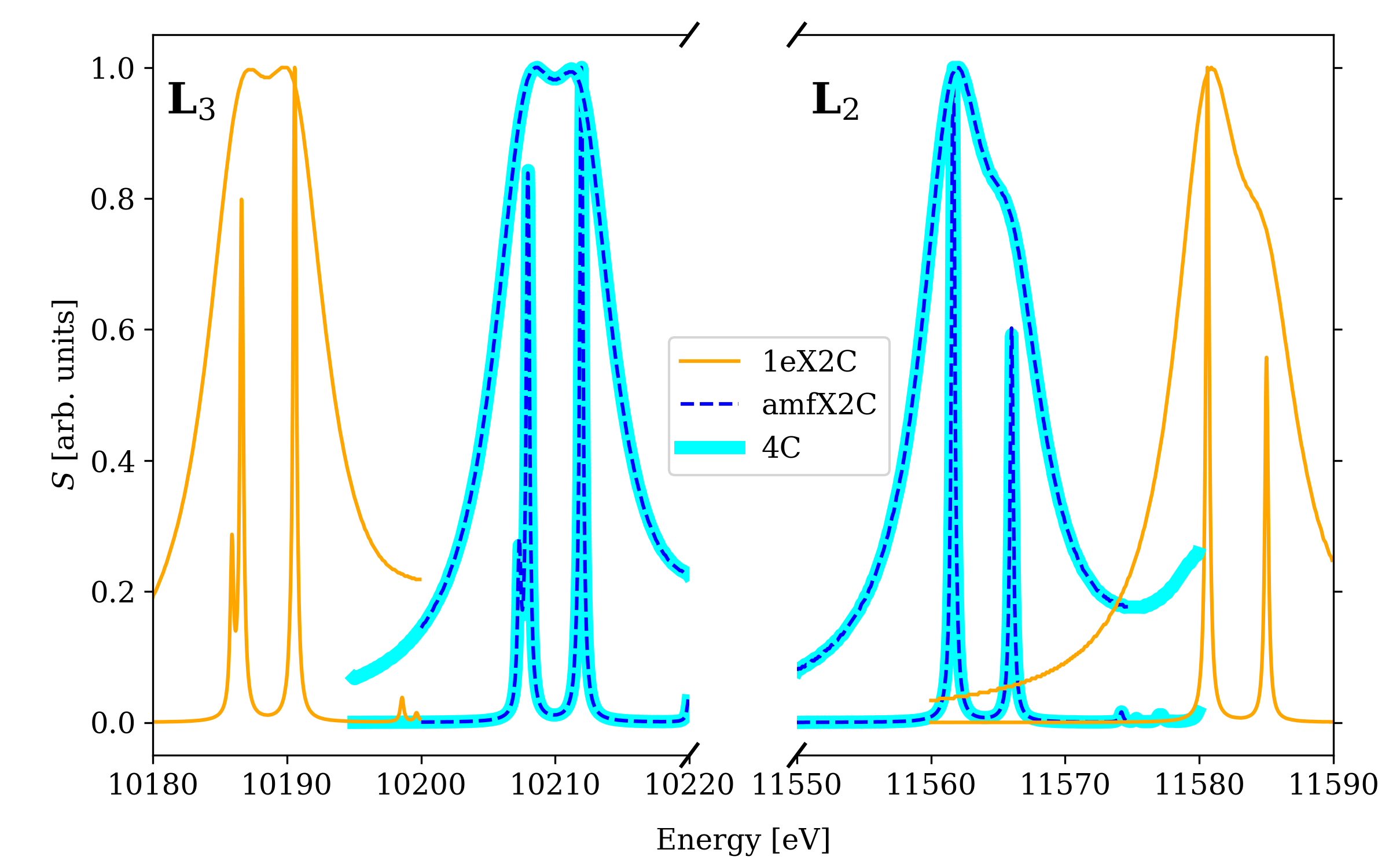}
\end{figure}

To determine the accuracy of the developed X2C DR-TDDFT approaches, we first repeated the calibration
study from our previous work focused on the 4c method.~\cite{Konecny2022}
The calibration set consists of XAS spectra near \ce{L_{2,3}}- and \ce{M_{4,5}}-edges of various
3d, 4d, 5d and 5f elements in small molecules.
For these we determined offsets from experimental values using 1eX2C, amfX2C and mmfX2C relativistic DR-TDDFT
and compared these offsets to the 4c results. The results are summarized in Table~\ref{tab:X2C}.
We see that the one-electron X2C variant (1eX2C) overestimates the spin--orbit splitting with respect
to the 4c calculations as well as to the experiment. This error can be attributed to not accounting
for the transformation of the two-electron term in the one-electron X2C approximation.
This shortcoming is remedied in the amfX2C and mmfX2C approaches, which exactly reproduce the 4c
spectral line positions. The agreement holds for the whole spectra as seen in
Figures~\ref{fig:MoS4-X2Cvs4C} and~\ref{fig:WCl6-X2C-SO} demonstrating the shortcomings
of 1eX2C and the improvements achieved by amfX2C and mmfX2C.
The eamfX2C method gives the same results as amfX2C as demonstrated in Table~S1 in
Section S1 in the Supporting information. However, in the rest of the paper, we focus
on the computationally simpler amfX2C approach.

While the overestimation of the SO splitting by 1eX2C is a \textit{quantitative} effect of shifting the
excitation energies, in calculations of the whole spectra, it can also be manifested as a \textit{qualitative}
change of spectral shapes. This happens particularly in cases when edges separated by SO splitting
(\ce{L2}-\ce{L3}, \ce{M2}-\ce{M3}, \ce{M4}-\ce{M5}) overlap, and peaks from both edges fall into
the same energy window. This effect is illustrated for the spectra near the molybdenum \ce{M4}- and \ce{M5}-edges
of  \ce{MoS4^{2-}} (PBE0 functional, DZ/aDZ basis sets) depicted in Figure~\ref{fig:MoS4-X2Cvs4C}.
While amfX2C reproduces the reference 4c results exactly, 1eX2C overestimates
the SO splitting causing a red shift of the \ce{M5} lines and blue shift of the \ce{M4} lines.
Namely, lines A, B, D belonging to the \ce{M5}-edge are in the 1eX2C description shifted to lower energies
(full arrows), while lines C and E belonging to the \ce{M4}-edge are shifted to higher energies
(empty arrows). As a consequence, lines C and D overlap and merge in the 1eX2C spectrum with the given
broadening parameter, resulting in a different overall spectral shape when compared to 4c and amfX2C data.

Since amfX2C, eamfX2C, and mmfX2C reproduce the 4c spectra, they allow the same computational protocol
developed in the context of 4c calculations to be reused with these 2c Hamiltonians.
The protocol aims to reproduce experimental spectral line positions to avoid arbitrary shifting
of calculated spectra and is based on increasing the admixture of Hartree--Fock exchange (HFX)
in hybrid functionals, with the optimal value being 60\% above \unit[1000]{eV} independent on the underlying
pure xc potential, e.g. PBE0 or B3LYP~\cite{Slater1951,Vosko1980,Becke1988,Lee1988,Stephens1994}
This is contrasted with 1eX2C, whose overestimation of the spin--orbit coupling is also present
in the calculations with increased HFX (see Figure~\ref{fig:WCl6-X2C-SO}). As a result, the spectra
calculated with the HFX amount determined at the 4c level of theory miss the experimental edge positions.
1eX2C would thus need its own computational protocol that would have to rely on error cancellation.
However, using amfX2C, eamfX2C, and mmfX2C, we can reproduce the experimental data at the two-component
relativistic level of theory in the same way as at the 4c level, see the last four
columns of Table~\ref{tab:X2C} and Table~S1.

A motivation for the development of two-component methods is their lower computational cost.
To determine the performance of the X2C DR-TDDFT in this regard, a comparison of computational
times required for a single iteration of DR-TDDFT calculations is reported in Table~\ref{tab:timings}
along with times required for SCF iterations
for the systems considered in the benchmark study as well as one of the larger complexes.
Here we consider DR-TDDFT calculations for a single frequency point rather than for 100 frequencies treated
together (the set-up used in the other calculations in this paper). This choice is made to allow for an equivalent
comparison between the 2c and 4c levels of theory, since the details of the multi-frequency implementations
in our program differ between these cases.
Therefore, these times serve only as a measure of the speed-up achieved by X2C DR-TDDFT against 4c DR-TDDFT rather than
of the total computational cost of the calculations.

An iteration of a subspace solver of relativistic X2C as well as 4c DR-TDDFT 
is dominated by the calculation of the two-electron integrals that are needed to
obtain the elements of matrices $\mathbf{A}^{\!\text{2c}}$ and $\mathbf{B}^{\!\text{2c}}$
in Eq.~\eqref{eq:linearDampedResponse}.
These matrix elements are on-the-fly contracted with the elements of the so-called trial vectors
that constitute the basis of the subspace in which the solution $(\mathbf{X}_u \mathbf{Y}_u)^\mathrm{T}$
is sought (see the algorithm in Ref.~\citenum{Konecny2019}).
Since the trial vectors are constructed with defined hermitian and time-reversal symmetries,
only those elements of matrices $\mathbf{A}^{\!\text{2c}}$ and $\mathbf{B}^{\!\text{2c}}$
giving non-zero contributions are calculated owing to the efficient quaternion implementation.
The calculations were performed on a single computer node equipped with an Intel Xeon-Gold 6138 2.0 GHz processor
with 40 CPU cores. The Intel ifort 18.0.3 compiler with -O2 optimization and the parallel Intel
MKL library were used for compilation and linking.
The times show an approximately 7-fold speed-up achieved by amfX2C across the systems,
which is similar to the performance reported in our earlier work for real-time 1eX2C TDDFT~\cite{Konecny2016},
while the speed-up in the SCF step is also in line with previous results.~\cite{ReSpect}
The acceleration achieved by amfX2C together with its accuracy thus pushes the boundaries of
relativistic XAS calculations towards larger and experimentally more relevant systems.

There are several factors affecting total calculation times such as the number of iterations
needed for convergence and the number of trial vectors generated in one iteration.
For a given molecular system and basis set, these depend on
i) the chosen frequency range -- spectrally dense regions require more trial vectors and iterations;
ii) the damping parameter $\gamma$ -- calculations with smaller $\gamma$ require more trial vectors and iterations;
iii) the method for isolating core excitations -- the technique based on zeroing elements of the
perturbation operator used in our previous work~\cite{Konecny2022} is more demanding than core--valence
separation used in this work, since additional (small-amplitude) elements of the response vectors need
to be converged as well.

With these caveats in mind, let us examine total calculation times required to obtain XAS spectra
of \ce{[WCl4(PMePh2)2]} utilizing PBE0-60HF functional and DZ/aDZ basis sets.
Here, we used the same hybrid-parallel computational setup as in our previous reference 4c calculations~\cite{Konecny2022},
utilizing 16 computer nodes each equipped with AMD Epyc 7742 2.25GHz processors with 128 CPU cores.
The hybrid parallelization facilitates OpenMPI library (version 4.0.3) with 8 MPI processes per node 
and 16 OMP threads per MPI process. The compilation was done using Intel ifort 19.1.1.217 compiler with -O2 optimization
and linked to the in-build OMP-parallel Intel MKL library. 
For the case of high density-of-states regions, namely \unit[11560-11570]{eV} (W \ce{L2}-edge) and
\unit[10205-10215]{eV} (W \ce{L3}-edge), each comprising 100 frequency points, the 4c calculations
with $\gamma=\unit[3.0]{eV}$ solving the full DR-TDDFT equation with elements of the perturbation
matrix outside the core--virtual orbital pairs set to zero,
lasted approximately 85h 49m and 88h 59m, respectively.
Equivalent amfX2C calculations requiring similar number of iterations ($\pm 1$) and trial vectors,
took approximately 7h 33m (speed-up 11.4) and 8h 36m (speed-up 10.3).
The use of CVS in amfX2C brought the CPU times down to 1h 39m and 2h 8m mainly by decreasing the
total number of iterations and trial vectors.
As a rule of thumb, we can therefore conclude that a calculation that would have previously taken a week
can now be finished in less than a day while utilizing the same computational resources without
the loss of accuracy.

\begin{sidewaystable}
\caption{Main line positions and spin--orbit splittings in XAS spectra calculated using
DR-TDDFT with PBE0 and PBE0-$x$HF functionals in Dyall-DZ/aDZ basis sets
at the X2C relativistic level of theory (1e, (e)amf, mmf) compared
with fully relativistic 4c results and experimental data. The amf and eamf
Hamiltonians give identical results, see Table~S1 in the Supporting Information.
}
\label{tab:X2C}
\begin{tabular}{ccrrrrrrrrrrrr}
\toprule
                  &            & Exp.$^{a}$ & & \multicolumn{4}{c}{PBE0-25HF}           & & \multicolumn{4}{c}{PBE0-$x$HF}                 \\ 
                  &            &            & &  4c      & 1eX2C   & (e)amfX2C & mmfX2C & & $x$HF &  4c      & 1eX2C   & (e)amfX2C & mmfX2C \\
  \cmidrule{3-3} \cmidrule{5-8} \cmidrule{10-14}
  \ce{VOCl3}      & \ce{L3}    & 516.9      & &  507.5   & 507.1   & 507.5   & 507.5     & & 50  &  515.2   &   514.9 &  515.2  & 515.2   \\
                  & \ce{L2}    & 523.8      & &  514.2   & 514.5   & 514.2   & 514.2     & & 50  &  522.0   &   522.3 &  522.0  & 522.0   \\[4pt]
                  & $\Delta$SO & 6.9        & &  6.7     &   7.4   &   6.7   &   6.7     & &     &  6.8     &     7.4 &    6.8  &   6.8   \\[7pt]
  \ce{CrO2Cl2}    & \ce{L3}    & 579.9      & &  571.4   & 571.0   & 571.4   & 571.4     & & 50  &  580.0   &   579.6 &  580.0  & 580.0   \\
                  & \ce{L2}    & 588.5      & &  579.5   & 579.8   & 579.5   & 579.5     & & 50  &  588.2   &   588.5 &  588.2  & 588.2   \\[4pt]
                  & $\Delta$SO &   8.6      & &    8.1   &   8.8   &   8.1   &   8.1     & &     &    8.2   &     8.9 &    8.2  &   8.2   \\[7pt]
  \ce{MoS4^{2-}}  & \ce{L3}    & 2521.7     & &  2489.3  & 2486.2  & 2489.3  &  2489.3   & & 60  &  2523.1  &  2517.8 & 2523.1  & 2523.1  \\
                  & \ce{L2}    & 2626.0     & &  2595.9  & 2597.8  & 2595.9  &  2595.9   & & 60  &  2627.6  &  2630.3 & 2627.6  & 2627.6  \\[4pt]
                  & $\Delta$SO & 104.3      & &  106.6   & 111.6   &  106.6  &  106.6    & &     &  104.5   &   112.5 &  104.5  &  104.5  \\[3pt]
                  & \ce{M5}    & 228.7      & &  225.4   & 225.0   &  225.4  &  225.4    & & 40  &  228.9   &   228.5 &  228.9  & 228.9   \\
                  & \ce{M4}    & 231.7      & &  228.7   & 229.3   &  228.7  &  228.7    & & 40  &  232.2   &   232.8 &  232.2  & 232.2   \\[4pt]
                  & $\Delta$SO & 3.0        & &  3.3     & 4.3     &    3.3  &  3.3      & &     &  3.3     &     4.3 &    3.3  & 3.3     \\[7pt]
  \ce{PdCl6^{2-}} & \ce{L3}    & 3177.8     & &  3138.2  & 3133.9  & 3138.2  &  3138.2   & & 60  &  3173.4  &  3169.3 & 3173.4  & 3173.4  \\
                  & \ce{L2}    & 3334.7     & &  3297.4  & 3300.1  & 3297.4  &  3297.4   & & 60  &  3332.9  &  3336.5 & 3332.9  & 3332.9  \\[4pt]
                  & $\Delta$SO & 156.9      & &  159.2   & 166.2   &  159.2  &  159.2    & &     &  159.5   &   167.2 &  159.5  & 159.5   \\[7pt]
  \ce{WCl6}       & \ce{L3}    & 10212.2    & &  10139.8 & 10117.5 & 10139.9 &  10139.8  & & 60  &  10207.3 & 10185.9 & 10207.3 & 10207.3 \\
                  & \ce{L2}    & 11547.0    & &  11492.8 & 11505.9 & 11492.7 &  11492.8  & & 60  &  11561.7 & 11580.6 & 11561.6 & 11561.7 \\[4pt]
                  & $\Delta$SO & 1334.8     & &  1353.0  & 1388.4  &  1352.8 &  1353.0   & &     &  1354.4  &  1394.7 &  1354.3 &  1354.4 \\[7pt]
  \ce{ReO4-}      & \ce{L3}    & 10542.0    & &  10472.0 & 10448.2 & 10471.7 &  10472.0  & & 60  &  10541.0 & 10518.6 & 10541.0 & 10541.0 \\
                  & \ce{L2}    &  ---       & &  11911.9 & 11925.9 & 11911.9 &  11911.9  & & 60  &  11982.1 & 12002.1 & 11982.1 & 11982.1 \\[4pt]
                  & $\Delta$SO &  ---       & &  1439.9  &  1477.7 &  1440.2 &   1439.9  & &     &  1441.1  &  1483.5 &  1441.1 &  1441.1 \\[7pt]
  \ce{UO2(NO3)2}  & \ce{M5}    &  ---       & &  3515.3  & 3504.5  & 3515.3  &  3515.3   & & 60  &  3549.9  &  3539.2 &  3549.9 & 3549.9  \\
                  & \ce{M4}    & 3727.0     & &  3693.2  & 3704.7  & 3693.2  &  3693.2   & & 60  &  3728.0  &  3740.5 &  3728.0 & 3728.0  \\
                  & $\Delta$SO &  ---       & &  177.9   &  200.2  &  177.9  &   177.9   & &     &  178.1   &   201.3 &   178.1 &  178.1  \\
\bottomrule
\end{tabular}

\vspace*{5pt}
$^{a}$ Experimental references:
\ce{VOCl3}: Ref.~\citenum{Fronzoni2009};
\ce{CrO2Cl2}: Ref.~\citenum{Fronzoni2009};
\ce{MoS4^{2-}}: Ref.~\citenum{George2009};
\ce{WCl6}: Ref.~\citenum{Jayarathne2014};
\ce{PdCl6^{2-}}: Ref.~\citenum{Boysen2008};
\ce{ReO4-}: Ref.~\citenum{Tougerti2012};
\ce{UO2(NO3)2}: Ref.~\citenum{Butorin2016}.
\end{sidewaystable}

\begin{table}[ht]
\caption{Computational cost of one iteration of SCF and DR-TDDFT calculations.\textsuperscript{\emph{a}}
}
\label{tab:timings}
\begin{tabular}{ccccrrrr}
\toprule
System               & &   Step         & & \multicolumn{2}{c}{Time}  & & Speed-up  \\
                     & &                & & 4c           &  amfX2C    & &           \\
\cmidrule{1-1} \cmidrule{3-3} \cmidrule{5-6} \cmidrule{8-8}
\ce{VOCl3}           & &    SCF         & &         9.2s &       2.6s & & 3.5       \\
                     & &    DR-TDDFT    & &        43.4s &       6.5s & & 6.7       \\
\ce{CrO2Cl2}         & &    SCF         & &         7.7s &       2.3s & & 3.3       \\
                     & &    DR-TDDFT    & &        37.9s &       5.8s & & 6.5       \\
\ce{MoS4^{2-}}       & &    SCF         & &        13.2s &       4.2s & & 3.1       \\
                     & &    DR-TDDFT    & &        52.4s &       8.4s & & 6.2       \\
\ce{PdCl6^{2-}}      & &    SCF         & &        29.6s &       9.3s & & 3.2       \\
                     & &    DR-TDDFT    & &     1m 46.7s &      16.7s & & 6.4       \\
\ce{WCl6}            & &    SCF         & &        59.2s &      17.2s & & 3.4       \\
                     & &    DR-TDDFT    & &     3m  0.3s &      26.4s & & 6.8       \\
\ce{ReO4-}           & &    SCF         & &        23.5s &       6.6s & & 3.6       \\
                     & &    DR-TDDFT    & &     1m 33.8s &      13.1s & & 7.2       \\
\ce{UO2(NO3)2}       & &    SCF         & &     2m 54.1s &      47.2s & & 3.7       \\
                     & &    DR-TDDFT    & &     8m 44.3s &      59.9s & & 8.8       \\
\ce{[WCl4(PMePh2)2]} & &    SCF         & & 1h 15m  2.2s &  24m 31.0s & & 3.1       \\
                     & &    DR-TDDFT    & & 4h 19m  1.3s &  37m 23.1s & & 6.9       \\
\bottomrule
\end{tabular}

\vspace{5pt}
\textsuperscript{\emph{a}}
The calculations were performed on a single computer node equipped with an Intel Xeon-Gold 6138 2.0 GHz processor
with 40 CPU cores. The Intel ifort 18.0.3 compiler with -O2 optimization and the parallel Intel
MKL library were used for compilation and linking.
\end{table}

\subsection{Larger systems}
\label{sec:Results:LargerSystems}

The main goal in the development of X2C-based DR-TDDFT is to allow multi-component relativistic
calculations to be applied to large systems of chemical interest, such as heavy metal-containing complexes with
complicated and heavy atom-containing ligands. In this section we report spectra of such systems.
Of the three X2C approaches, amfX2C, eamfX2C, and mmfX2C, that reproduced the 4c data in the calibration
presented in the previous section, we focus in the following on amfX2C.
This is because in amfX2C, the whole calculation including the initial ground-state SCF is performed
in a two-component regime and in a simpler way than in eamfX2C.
That is why we envision amfX2C to become the standard method of choice in future relativistic
calculations of XAS spectra of large molecular systems. 

\begin{figure}[ht]
\caption{
XAS spectrum of \ce{[Ru Cl2 (DMSO)2 (Im)2]} near overlapping Cl K-edge and Ru \ce{L3}-edge
calculated using PBE0-60HF functional and DZ/aDZ basis sets.
Both absorption edges are reproduced sufficiently well when different inverse lifetimes
($\gamma$) of core-excited states of different atoms are accounted for. 
No shift was applied on the energy axis to manually align the spectra.
}
\centering
\begin{subfigure}{0.29\textwidth}
  \includegraphics[width=0.95\textwidth]{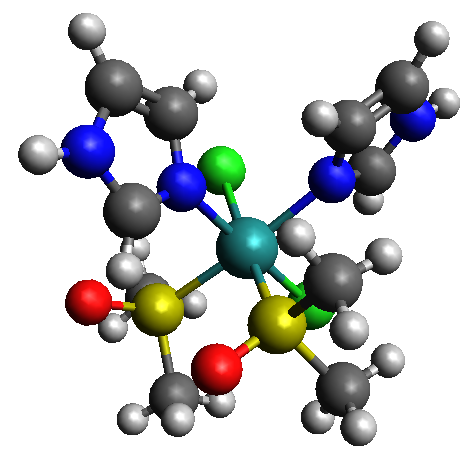}
  \caption{Structure}
  \label{fig:Ru_molecule}
\end{subfigure}
\begin{subfigure}{0.69\textwidth}
  \includegraphics[width=0.95\textwidth]{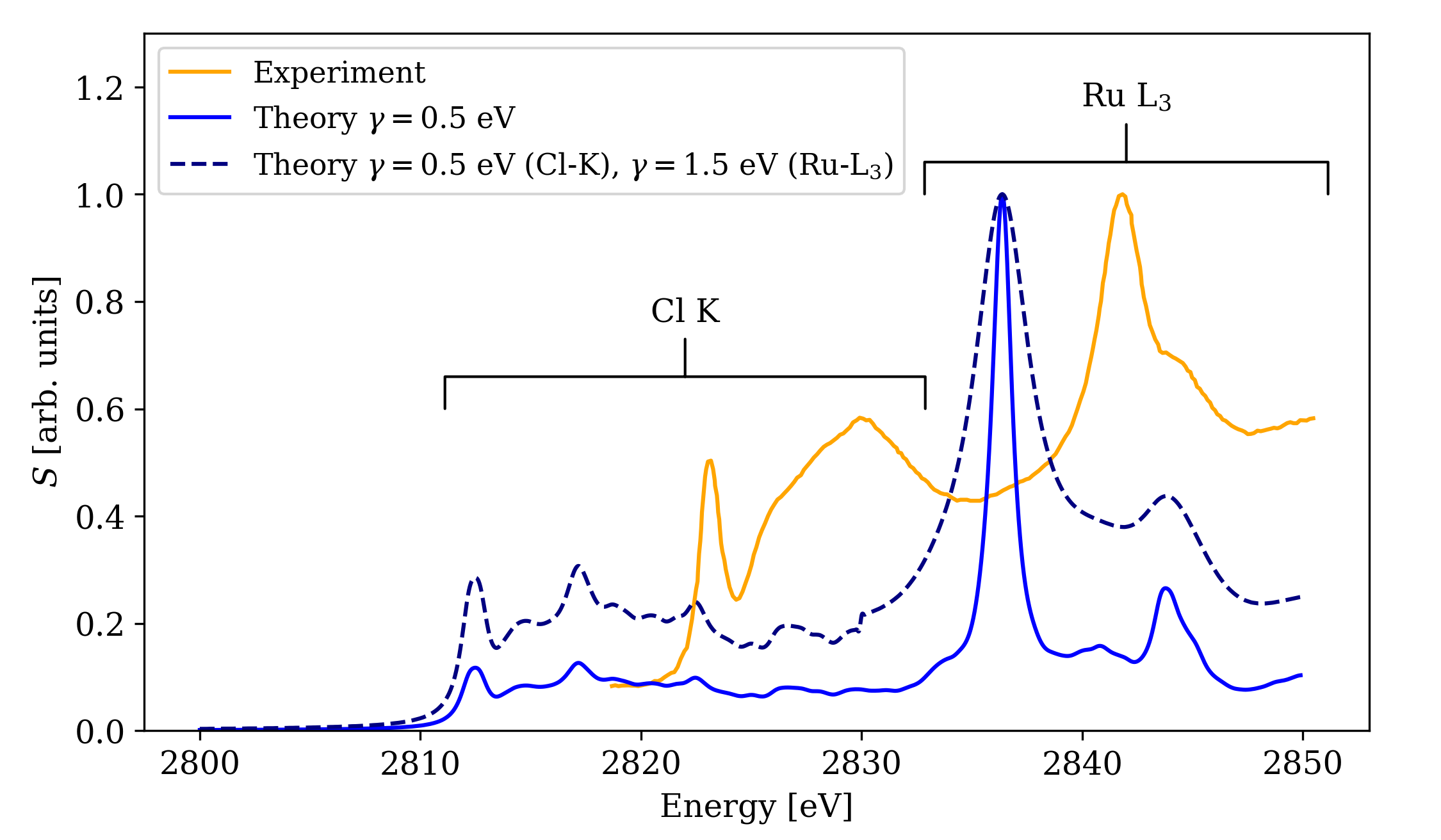}
  \caption{Spectrum}
  \label{fig:Ru_vsExp}
\end{subfigure}
\end{figure}

First, for \ce{[Ru Cl2 (DMSO)2 (Im)2]} (Figure~\ref{fig:Ru_molecule}) we calculated the spectra
in the region \unit[2800-2850]{eV}, covering both Cl K-edge as well as Ru \ce{L3}-edge
(Figure~\ref{fig:Ru_vsExp}).
The spectra are in general well aligned with experiment with differences of \unit[11]{eV}
and \unit[5.5]{eV} in the Cl K-edge and Ru \ce{L3}-edge positions, respectively, corresponding
to a slight, \unit[5.5]{eV}, overestimation of the separation between the edges.
In addition, as noted before~\cite{Konecny2022}, the position of lines within the same absorption
edge increases with the amount of HFX in the functional, resulting in a somewhat wider \ce{L3}-edge peak
in the calculation.
While we previously reported spectra calculated with a single damping parameter $\gamma=\unit[0.5]{eV}$
for the whole spectral range, here we also performed calculations with a larger damping parameter
$\gamma=\unit[1.5]{eV}$ in the region near the Ru \ce{L3}-edge, corresponding to a shorter lifetime of
the Ru \ce{p_{3/2}}-excited state. Since this damping parameter was determined from the experimental
spectrum, it led to a better agreement with the experimental reference.
This supports an idea also suggested in our previous work~\cite{Konecny2022} for a computational
protocol utilizing different damping parameters in CPP calculations of XAS spectra near overlapping
edges.

\begin{figure}[ht]
\caption{
amfX2C Hamiltonian reproduces the 4c reference while 1eX2C Hamiltonian overestimates
spin--orbit splitting leading to \unit[20]{eV} shifts 
in XAS spectra of \ce{[WCl4(PMePh2)2]} near W \ce{L_{2,3}}-edges calculated using PBE0-60HF functional
and DZ/aDZ basis sets.  Broad peaks were obtained with damping parameter $\gamma=\unit[3.0]{eV}$
while narrow ones with $\gamma=\unit[0.15]{eV}$.
}
\label{fig:W_complex-X2C-SO}
\centering
\includegraphics[width=0.65\textwidth]{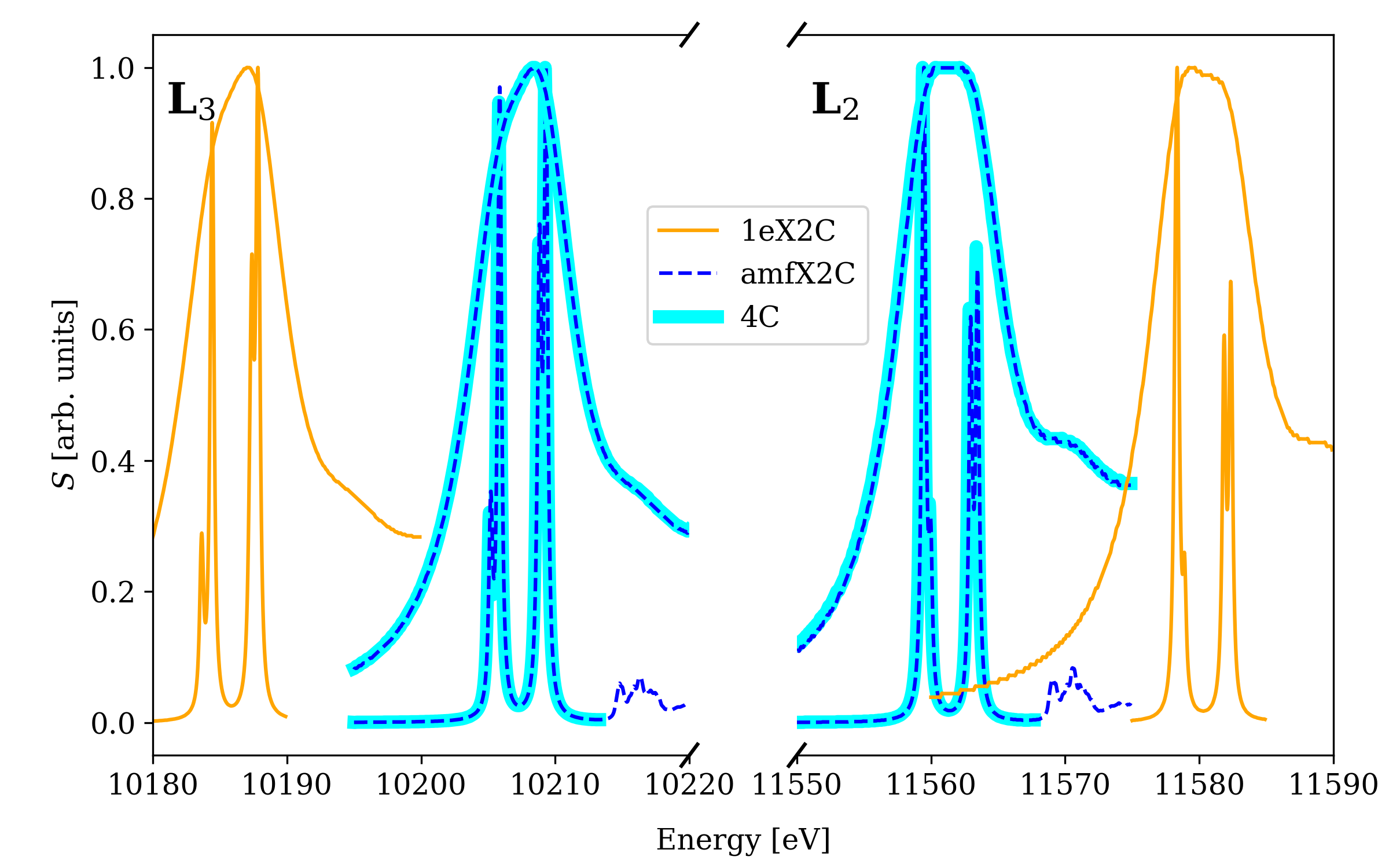}
\end{figure}

\begin{figure}[ht]
\caption{
Comparison of damped response (DR) and eigenvalue (EV) TDDFT
for XAS spectra of \ce{[WCl4(PMePh2)2]} near W \ce{L_{2,3}}-edges calculated using PBE0-60HF functional
and DZ/aDZ basis sets.
}
\centering
\begin{subfigure}{0.29\textwidth}
  \includegraphics[width=0.95\textwidth]{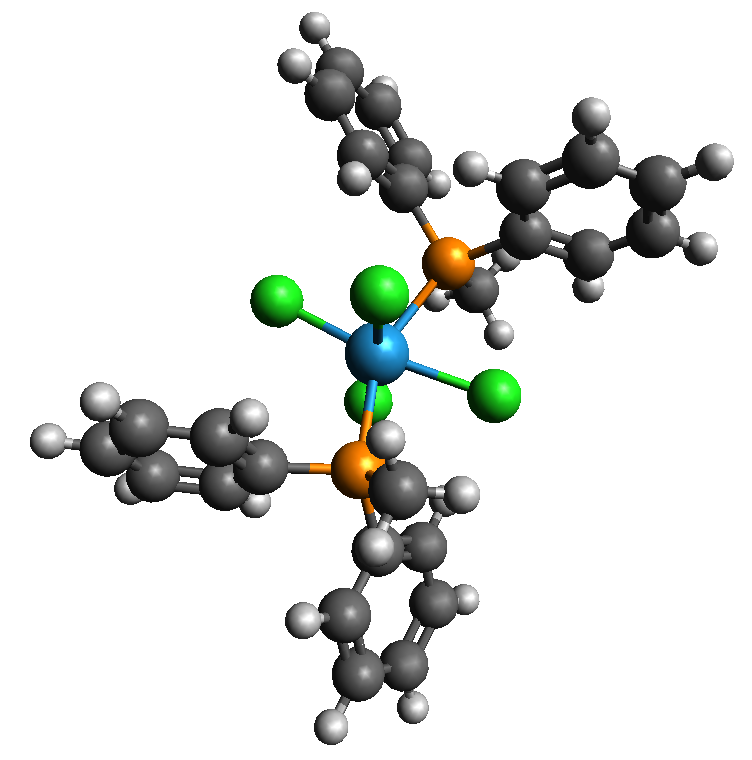}
  \caption{Structure}
  \label{fig:W_molecule}
\end{subfigure}
\begin{subfigure}{0.69\textwidth}
  \includegraphics[width=0.95\textwidth]{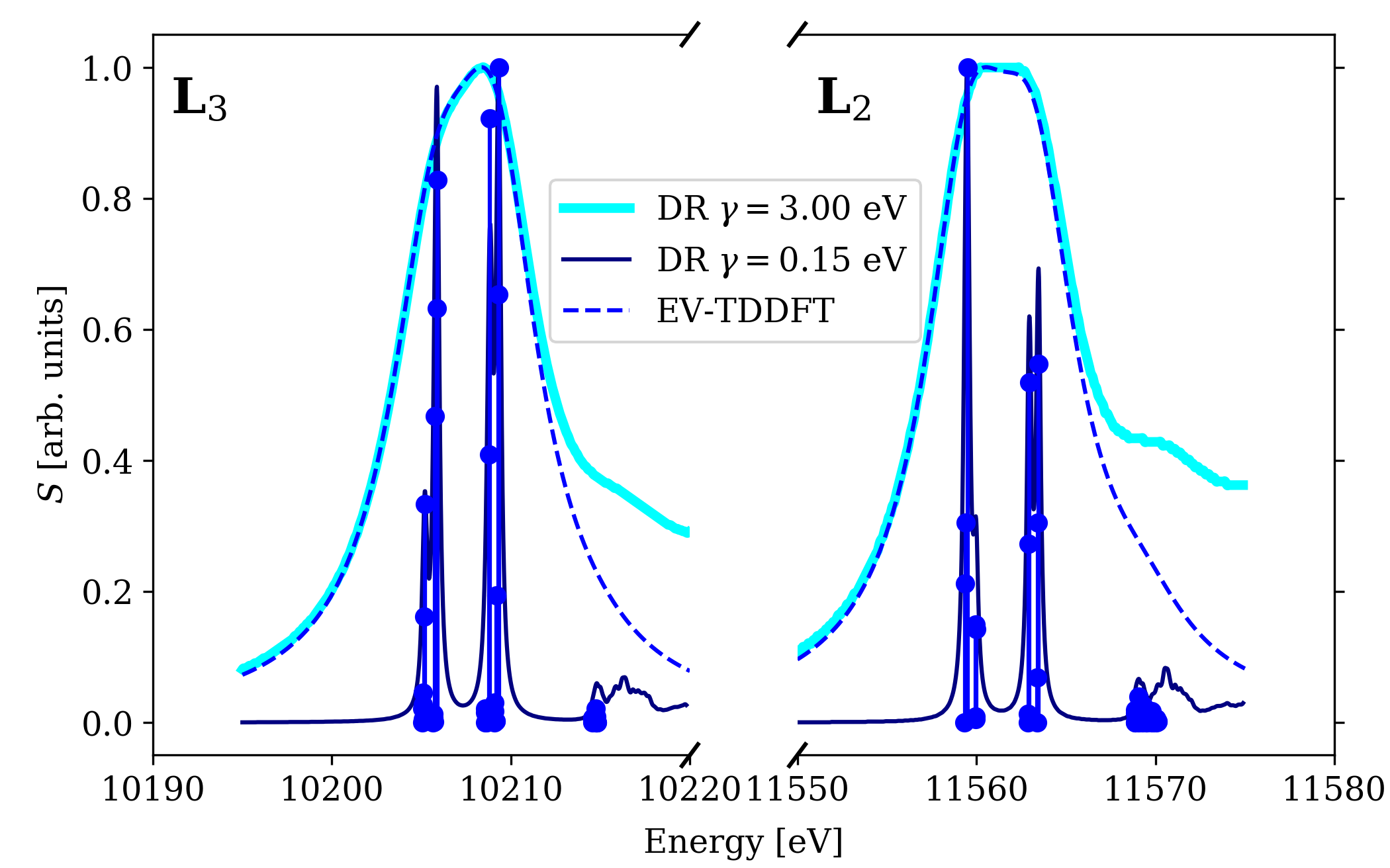}
  \caption{Spectrum}
  \label{fig:W_vsExp}
\end{subfigure}
\end{figure}

Second, we calculate XAS spectra near tungsten \ce{L_{2,3}}-edges of \ce{[W Cl4 (PMePh2)2]} (Figure~\ref{fig:W_molecule})
at the amfX2C relativistic level of theory. In addition to reproducing the 4c DR-TDDFT results,
we also document the failure of 1eX2C in Figure~\ref{fig:W_complex-X2C-SO} similarly as
in Figure~\ref{fig:WCl6-X2C-SO} for \ce{WCl6}. The conclusion is the same: 1eX2C is unable to
reproduce the reference 4c results due to its overestimation of spin--orbit splitting.
Moreover, we include the calculation of excitation energies and transition moments using eigenvalue TDDFT.
The final spectra are shown in Figure~\ref{fig:W_vsExp} and the calculated eigenvalues and corresponding
oscillator strengths are reported in the Supporting Information, Section S3, Tables S2 and S3.
The comparison of DR-TDDFT and EV-TDDFT showcases the pros and cons of these linear response TDDFT approaches.
On the one hand, DR-TDDFT gives the full spectral function on the frequency interval of interest
for a given damping parameter $\gamma$, which allows the experimental spectra to be reproduced.
However, in order to resolve the broad peaks into individual transitions, one has to decrease the
damping parameter. While this can yield fruitful results and interpretation of spectra
(see our analysis of this system at the 4c level of theory in Ref.~\citenum{Konecny2022}),
the EV-TDDFT accesses individual transitions directly, essentially in the limit $\gamma \rightarrow 0$.
On the other hand, in the iterative solution of EV-TDDFT, the user-defined number of eigenvalues is calculated
from the lowest for the edge specified by the core--valence separation. This explains why the two methods
initially lead to the same spectral function but start to depart for higher energies: more than the 50
transitions considered in EV-TDDFT would have been needed to match the spectra. The higher number of
eigenvalues increases the computational cost of the method as well as puts strains on the stability
of the iterative solver. The choice of the best-suited method thus depends on the chemical problem at hand
and this example showcases that the amfX2C-based implementation of both DR- and EV-TDDFT in the \ReSpect{}
program is up to the task of calculating XAS spectra of large complexes with heavy metal central atoms.

\begin{figure}[ht]
\caption{
Comparison of experimental and theoretical damped response (DR) and eigenvalue (EV) TDDFT
XAS spectra of \ce{[(\eta^{6}-p-cym)Os(Azpy-NMe2)I]^{+}} near Os \ce{L_{3}}-edge
calculated using PBE0-60HF functional and DZ/aDZ basis sets.
No shift was applied on the energy axis to manually align the spectra.
}
\begin{subfigure}{0.29\textwidth}
  \includegraphics[width=0.95\textwidth]{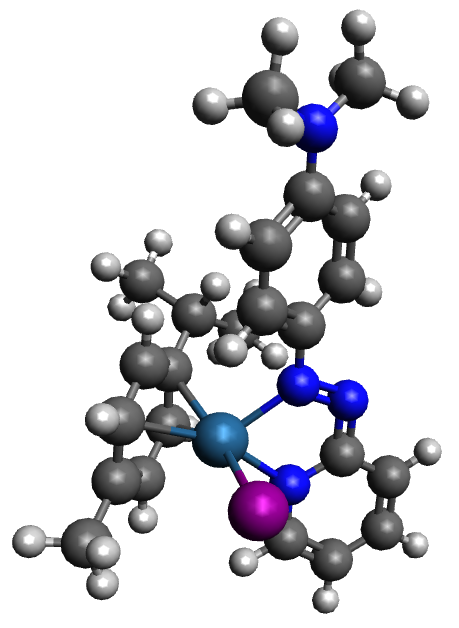}
  \caption{Structure}
  \label{fig:Os_molecule}
\end{subfigure}
\begin{subfigure}{0.69\textwidth}
  \includegraphics[width=0.95\textwidth]{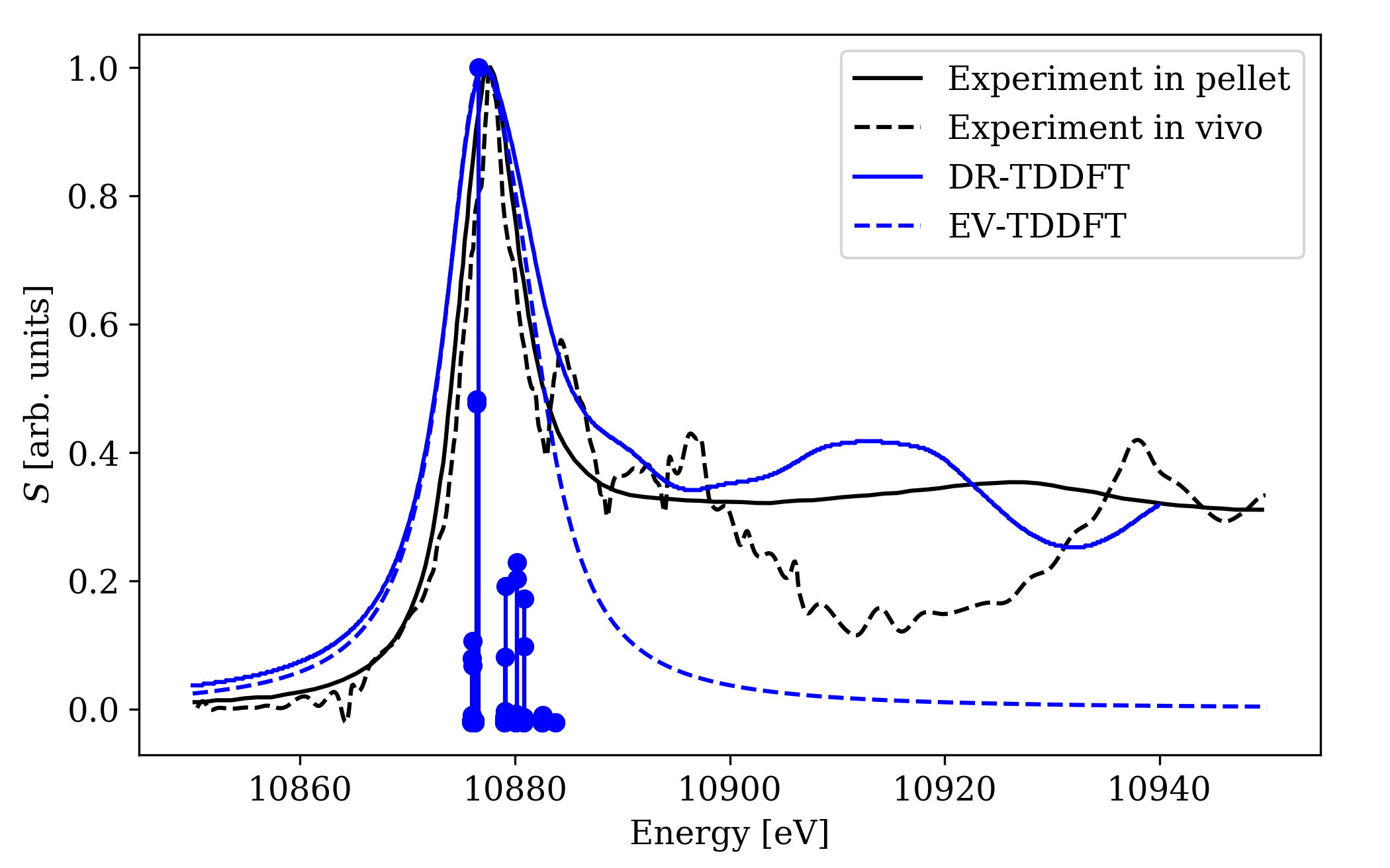}
  \caption{Spectrum}
  \label{fig:Os_vsExp}
\end{subfigure}
\end{figure}

In the final application we focus on a system not considered in our previous work,
\ce{[(\eta^{6}-p-cym)Os(Azpy-NMe2)I]^{+}}
that also contains the heavy iodine atom as a ligand in addition to the large organic ligands,
see Figure~\ref{fig:Os_molecule}.
The final spectra are shown in Figure \ref{fig:Os_vsExp} and the calculated eigenvalues and corresponding
oscillator strengths are reported in the Supporting Information, Section S3, Table S4.
The system was investigated experimentally by Sanchez-Cano et al.~\cite{SanchezCano2018}
as an anti-cancer drug and its XAS spectra near the Os \ce{L3}-edge were recorded both
inside a cellulose pellet as well as in a cell culture. Both spectra are dominated by
a major peak centred at \unit[10878]{eV}. The main features of the spectrum are reproduced
in the amfX2C calculation for the molecule \textit{in vacuo}. However, the reproduction of the
satellite signals would require a detailed study of environment effects which is beyond the scope
of the present work.
In this example, the combination of relativistic level of theory and the reparametrized PBE0-60HF
xc potential achieved a staggering alignment of the spectra on the energy axis where
\textit{no additional shift was applied} to manually align the spectra.

\section{Conclusions}
\label{sec:Conclusion}

We have presented a detailed theory derivation of X2C-based damped response time-dependent
density functional theory (DR-TDDFT) and eigenvalue TDDFT (EV-TDDFT) starting from the four-component (4c) time-dependent
Dirac--Kohn--Sham equations. We showed that X2C models known from time-independent
calculations including 1eX2C as well as mmfX2C and the recently introduced (e)amfX2C
can be extended to the time domain, and we derived the time-dependent Fock matrices
that for (e)amfX2C include the important two-electron picture-change effects.
We showed how initially time- and external-field dependent X2C transformation
matrices can be considered static for a weak field in the dipole approximation, which
allowed us to formulate linear response TDDFT in the damped response and eigenvalue formalisms,
where the final equations in molecular orbital basis
have the same form and properties as the 4c equations. This in turn enabled a straightforward
extension of the solvers previously developed at the 4c level of theory to these new
two-component Hamiltonians.

We presented benchmark results for XAS spectra of transition metal and actinide compounds
at the metal L- and M-edges, where spin--orbit (SO) splitting dominates the spectra. While the
1eX2C method overestimated the SO splitting, the mmfX2C and (e)amfX2C shined. The latter X2C
approaches fully reproduced the reference 4c results with a considerable saving of CPU time.
This agreement allowed us to use the computational protocol for XAS calculations optimized
at the 4c level of theory to be reused in X2C calculations. Since the assumption in 4c calculations
was that relativistic effects were solved by such a high-level relativistic treatment, the
same can be said about mmfX2C and (e)amfX2C calculations. This is contrasted with 1eX2C where a new computational
protocol relying on error cancellation would have to be developed.

As final highlights we presented calculated XAS spectra of large systems involving about 50 atoms:
\ce{[Ru Cl2 (DMSO)2 (Im)2]} (Im = imidazole, DMSO = dimethyl
sulfoxide), \ce{[W Cl4 (PMePh2)2]} (Ph = phenyl), and \ce{[(\eta^{6}-p-cym)Os(Azpy-NMe2)I]^{+}}
(p-cym = p-cymene, \ce{Azpy-NMe 2} = 2-(p-([dimethylamino]phenylazo)pyridine))),
and obtained an excellent agreement with experimental reference data both in terms of excitation
energies as well as line shapes.

Based on the presented results, demonstrating both accuracy and computational efficiency, we envision the
amfX2C and eamfX2C DR- and EV-TDDFT, where the full calculations including the initial SCF are
performed in a two-component regime, to become a new paradigm for relativistic calculations
of XAS spectra of large molecules containing heavy elements.

\section{Supporting information}

Molecular geometries, EV-TDDFT, and eamfX2C DR-TDDFT results.

\clearpage



\begin{acknowledgement}
We acknowledge the support received from the Research Council of
Norway through a Centre of Excellence Grant (No. 262695),
Research Grant (No. 315822), and a Mobility grant (No. 314814) as well as the use of 
computational resources provided by UNINETT Sigma2 -- The National
Infrastructure for High Performance Computing and Data Storage
in Norway (Grant No. NN4654K). In addition, this project received 
funding from the European Union’s Horizon 2020 research and innovation 
program under the Marie Skłodowska-Curie Grant Agreement No. 945478 (SASPRO2),
and the Slovak Research and Development Agency (Grant Nos. APVV-21-0497 and APVV-19-0516).
S.K. acknowledges the financial support provided by the Slovak Grant 
Agency VEGA (Contract No. 2/0135/21).
J.V. acknowledges the support of the Ministry of Education, Youth and Sports
of the Czech Republic project DKRVO (RP/CPS/2022/007).
\end{acknowledgement}

%
%
\begin{appendices}

\section{Limiting behavior of the X2C unitary transformation}
\label{appendixA}

In this section we prove the formulas for estimation of the time and field
derivative of the X2C unitary transformation when the system is in the presence of
an external oscillating field of frequency $\omega$ and amplitude
$|\boldsymbol{\mathcal{E}}|$.  The estimation expressions can be written as
\begin{alignat}{2}
  \label{eq:A:Udot}
  \mathbf{\dot{U}}(t,\boldsymbol{\mathcal{E}})
  &=
  O(|\boldsymbol{\mathcal{E}}|\omega c^{-1})
  \qquad
  &&\text{as} \,\,c \rightarrow \infty
  \quad \wedge \quad |\boldsymbol{\mathcal{E}}| \rightarrow 0,
  \\
  \label{eq:A:Uprime}
  \mathbf{U}^{(1)}(0,\boldsymbol{\mathcal{E}})
  &=
  O(|\boldsymbol{\mathcal{E}}| c^{-1})
  \qquad
  &&\text{as} \,\,c \rightarrow \infty,
\end{alignat}
where we used a notation that distinguishes between field-dependent,
$x^{(n)}(t)$, and field-independent, $x^{(n)}_u(t)$, variables as defined
in the following perturbational expression
\begin{gather}
  \label{eq:def-response}
  x(t,\boldsymbol{\mathcal{E}}) = x^{(0)}(t) + x^{(1)}(t) + \cdots,
  \\
  x^{(0)}(t) \coloneqq x(t,0),
  \qquad
  x^{(1)}(t) = \sum^3_u \mathcal{E}_u x_u^{(1)}(t)
  \coloneqq \sum^3_u \mathcal{E}_u
    \left.
    \frac{\partial x(t,\boldsymbol{\mathcal{E}})}{\partial \mathcal{E}_u}
    \right|_{\boldsymbol{\mathcal{E}}=0}.
\end{gather}

In the subsequent discussion it becomes advantageous to expand the X2C unitary
transformation\cite{Konecny2016}
\begin{gather}
  \mathbf{U}(t,\boldsymbol{\mathcal{E}}) = 
  \begin{pmatrix}
      \mathbf{I} & -\mathbf{R}^\dagger \\
      \mathbf{R} & \mathbf{I}
  \end{pmatrix}
  \begin{pmatrix}
      (\mathbf{I} + \mathbf{R}^\dagger\mathbf{R})^{-1/2} & \mathbf{0} \\
      \mathbf{0} & (\mathbf{I} + \mathbf{R}\mathbf{R}^\dagger)^{-1/2}
  \end{pmatrix},
\end{gather}
in its Taylor series around $\mathbf{R} = 0$
\begin{equation}
  \label{eq:A:Uexpand}
  \mathbf{U}(t,\boldsymbol{\mathcal{E}})
  =
  \begin{pmatrix}
      \mathbf{I} & \mathbf{0} \\
      \mathbf{0} & \mathbf{I}
  \end{pmatrix}
  +
  \begin{pmatrix}
      \mathbf{0} & -\mathbf{R}^{\dagger} \\
      \mathbf{R} & \mathbf{0}
  \end{pmatrix}
  +
  O(\|\mathbf{R}\|^2)
  \qquad
  \text{as} \,\,\|\mathbf{R}\| \rightarrow 0,
\end{equation}
where the brackets, $\|.\|$, represent some suitable matrix norm, {\it e.g.}
Frobenius norm.  Here and in the following we employ a simplified notation for
the coupling matrix, $\mathbf{R} \coloneqq
\mathbf{R}(t,\boldsymbol{\mathcal{E}})$.  From Eq.~\eqref{eq:A:Uexpand} it is
clear that to estimate the linear response of the matrix
$\mathbf{U}(t,\boldsymbol{\mathcal{E}})$, it is sufficient to analyze the behavior
of the coupling matrix $\mathbf{R}$. To obtain the
limiting behavior of $\mathbf{R}$, one may explore its dependence on
time, field, and speed of light $c$.

In the next paragraph, we analyze the limiting behavior of $\mathbf{R}$ with
respect to $c$.  For that purpose, we may utilize the definition of the X2C
unitary transformation presented in Eq.~\eqref{decoupled_F_C}, where the
off-diagonal---LS and SL---blocks of the transformed matrix are set to zero.
However, a more transparent alternative is to use the expression for determining
the coupling matrix within the one-step X2C method, see Eq.~(23) in
Ref.~\citenum{Ilias2007} or Eq.~(8) in Ref.~\citenum{Konecny2016}
\begin{gather}
  \label{eq:A:Rformula}
    \mathbf{C}^\mathrm{S}_\mathrm{+}
  - \mathbf{R} \mathbf{C}^\mathrm{L}_\mathrm{+}
  \overset{!}{=} 0,
\end{gather}
with $\mathbf{C}_\mathrm{+}$ being the 4c positive-energy MO coefficients.  If
one uses orthonormal basis, then $\mathbf{C}^\mathrm{L}_\mathrm{+}$ and
$\mathbf{C}^\mathrm{S}_\mathrm{+}$ are of order $c^0$ and $c^{-1}$,
respectively.  If we assume that the left-hand-side of
Eq.~\eqref{eq:A:Rformula} has a convergent Laurent series with $c$ as the
variable, then the condition, $\dots\overset{!}{=} 0$, is satisfied for each
order of this series separately.  By analyzing these conditions order by order
it follows that the first nonzero term in the expansion of the coupling matrix
is of order $c^{-1}$ and therefore its limiting behavior can be written as
\begin{gather}
  \label{eq:A:Rc}
  \mathbf{R} = O(c^{-1})
  \qquad
  \text{as} \,\,c \rightarrow \infty.
\end{gather}

To further improve the estimation of the coupling matrix, one can explore its
time and field dependence.  If the molecular system is subject to the
time-dependent external electric field, $\mathbf{E}(t)$, the time-dependent
Kohn-Sham equation can be viewed as a representation of a continuous
time-invariant system that assigns the output signal---induced electric dipole
moment, X2C unitary transformation, and other time-dependent quantities---to
the input signal, $\mathbf{E}(t)$.  By choosing the harmonic time-dependence of
the electric field, $\mathbf{E}(t) = \boldsymbol{\mathcal{E}}
\mathrm{cos}(\omega t)$, the coupling matrix $\mathbf{R}$, its time derivative,
and its linear response, can be expressed using the Volterra series as
follows
\begin{alignat}{2}
  \label{eq:A:RTaylor}
  \mathbf{R}(t,\boldsymbol{\mathcal{E}})
  &= \mathbf{R}(0,0)
  + \boldsymbol{\mathcal{R}}_u(\omega) \,\mathrm{cos}(\omega t) \,\mathcal{E}_u
  + O(|\boldsymbol{\mathcal{E}}|^2)
  \qquad
  &&\text{as} \,\,|\boldsymbol{\mathcal{E}}| \rightarrow 0,
  \\
  \label{eq:A:Rdot}
  \mathbf{\dot{R}}(t,\boldsymbol{\mathcal{E}})
  &= -\omega\, \boldsymbol{\mathcal{R}}_u(\omega) \,\mathrm{sin}(\omega t) \,\mathcal{E}_u
  + O(|\boldsymbol{\mathcal{E}}|^2)
  \qquad
  &&\text{as} \,\,|\boldsymbol{\mathcal{E}}| \rightarrow 0,
  \\
  \label{eq:A:Rprime}
  \mathbf{R}^{(1)}(t,\boldsymbol{\mathcal{E}})
  &= \boldsymbol{\mathcal{R}}_u(\omega) \,\mathrm{cos}(\omega t) \,\mathcal{E}_u.
\end{alignat}
Combining Eqs.~\eqref{eq:A:Rc}--\eqref{eq:A:Rprime} one can write the
limiting behavior of the coupling matrix as
\begin{alignat}{2}
  \label{eq:A:Rdot1}
  \mathbf{\dot{R}}(t,\boldsymbol{\mathcal{E}})
  &=
  O(|\boldsymbol{\mathcal{E}}|\omega c^{-1})
  \qquad
  &&\text{as} \,\,c \rightarrow \infty
  \quad \wedge \quad |\boldsymbol{\mathcal{E}}| \rightarrow 0,
  \\
  \label{eq:A:Rprime1}
  \mathbf{R}^{(1)}(t,\boldsymbol{\mathcal{E}})
  &=
  O(|\boldsymbol{\mathcal{E}}| c^{-1})
  \qquad
  &&\text{as} \,\,c \rightarrow \infty.
\end{alignat}
Finally, to prove Eqs.~\eqref{eq:A:Udot} and~\eqref{eq:A:Uprime}, one simply
uses the linear part of the expansion in Eq.~\eqref{eq:A:Uexpand} and the
estimations for the coupling matrix $\mathbf{R}$ in Eqs.~\eqref{eq:A:Rdot1}
and~\eqref{eq:A:Rprime1}.  {\it Q.E.D.}

%
%
\section{Contribution of the negative-energy states to 2c EOM}
\label{appendixB}

The transformation of the four-component EOM, Eq.~\eqref{4c:eom}, by the X2C
unitary matrix $\mathbf{U}$ results in two separate two-component equations
[see Eq.~\eqref{x4c:eom:approx}].  However, these equations are still coupled,
because, for example, the LL equation whose solutions are
$\boldsymbol{\tilde{C}}^\mathrm{L}_{i} (t,\boldsymbol{\mathcal{E}})$ depends on
the solutions of the SS equation, $\boldsymbol{\tilde{C}}^\mathrm{S}_{i}
(t,\boldsymbol{\mathcal{E}})$, through the density matrix
$\mathbf{\tilde{D}}^\mathrm{4c}(t,\boldsymbol{\mathcal{E}})$.  In addition, the
complex linear electric dipole polarizability tensor
$\boldsymbol{\alpha}(\omega)$, Eq.~\eqref{eq:polarisability-DR}, whose imaginary
part determines the electronic absorption spectrum
[Eq.~\eqref{eq:rspFunction}], depends on both solutions
$\boldsymbol{\tilde{C}}^\mathrm{L}_{i} (t,\boldsymbol{\mathcal{E}})$ and
$\boldsymbol{\tilde{C}}^\mathrm{S}_{i} (t,\boldsymbol{\mathcal{E}})$.
Therefore, in contrast to the static case, in the time domain it does not suffice
to fully transform 4c EOM to one 2c EOM  to find the X2C unitary transformation,
and one rather needs to solve two 2c EOMs.  However, thanks to the X2C
unitary transformation, the large component (small component) of the time- and
field-dependent occupied MO coefficients,
$\boldsymbol{\tilde{C}}^{\mathrm{4c}}_i(t,\boldsymbol{\mathcal{E}})$, depend
only on the positive-energy (negative-energy) reference MOs,
$\boldsymbol{\tilde{C}}_{+}$ ($\boldsymbol{\tilde{C}}_{-}$) [see
Eq.~\eqref{x4c:ansatz-mmf}]
\begin{equation}
  \boldsymbol{\tilde{C}}_{+}
  \coloneqq
  \mathbf{U}^\dagger {\boldsymbol{C}}^\mathrm{4c}_{+}
  =
  \left(
    \begin{array}[c]{c}
      \boldsymbol{\tilde{C}}^\mathrm{L}_{+}  \\
      \boldsymbol{0}                            \\
    \end{array}
  \right),
  \qquad
  \boldsymbol{\tilde{C}}_{-}
  \coloneqq
  \mathbf{U}^\dagger {\boldsymbol{C}}^\mathrm{4c}_{-}
  =
  \left(
    \begin{array}[c]{c}
      \boldsymbol{0}                            \\
      \boldsymbol{\tilde{C}}^\mathrm{S}_{-}  \\
    \end{array}
  \right).
\end{equation}
The contribution of the negative-energy states to the complex
polarizability tensor $\boldsymbol{\alpha}$ is only of order $c^{-4}$, and
can thus safely be neglected. Eq.~\eqref{x4c:eom:approx} is then fully decoupled
and it becomes sufficient to solve only one 2c EOM, Eq.~\eqref{x2c:eom} to
obtain the complex polarizability tensor and resulting electronic absorption
spectrum.

To prove that the effect of the negative-energy states in the complex
polarizability tensor is of order $c^{-4}$, one can simply examine the
expressions for the tensor $\boldsymbol{\alpha}$,
Eq.~\eqref{eq:polarisability-DR}, and DR-TDDFT
equations, Eq.~\eqref{eq:linearDampedResponse}. Note that we get the same result
regardless of whether we use the two- or four-component picture for the proof. One can then
employ a bit simpler four-component equations, where in the orthonormal basis
it holds that
\begin{gather}
  {\boldsymbol{\tilde{C}}}^\mathrm{4c}_{+} = O(c^{0})
  \quad
  \wedge
  \quad
  {\boldsymbol{\tilde{C}}}^\mathrm{4c}_{-} = O(c^{-1})
  \quad
  \wedge
  \quad
  \varepsilon_+ = O(c^{0})
  \quad
  \wedge
  \quad
  \varepsilon_- = O(c^{2})
  \\
  \quad
  \Rightarrow
  \quad
  P_{-+} = \Omega_{-+} = O(c^{-1})
  \quad
  \wedge
  \quad
  \Omega_{++} = O(c^{0})
  \quad
  \wedge
  \quad
  \Omega_{--} = O(c^{-2})
\end{gather}
as $c \rightarrow \infty$. Based on the DR-TDDFT
equation, Eq.~\eqref{eq:linearDampedResponse} one can then estimate the response
coefficients for the negative-energy states as, $X_{-+} = Y_{-+} = O(c^{-3})$,
and their contribution to the complex polarizability
tensor, Eq.~\eqref{eq:polarisability-DR} as $O(c^{-4})$.  {\it Q.E.D.}

\end{appendices}


%
%

\bibliography{bibliography}

\newpage
\begin{figure}
\label{TOC Graphic}
\caption*{TOC Graphic}
\includegraphics[width=3.25in]{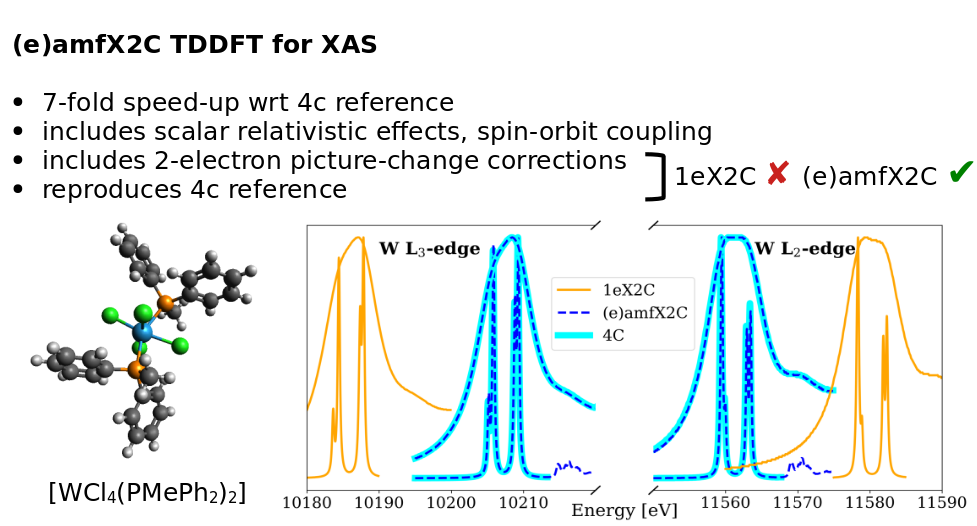}
\end{figure}

\end{document}